# Toward a benchmark for CTR prediction in online advertising: datasets, evaluation protocols and perspectives

Shan Gao[1], Yanwu Yang[1]

[1]School of Management, Huazhong University of Science and Technology, Wuhan 43004, China,
{gaoshan.isec, yangyanwu.isec}@gmail.com

**Abstract**: Click-through rate (CTR) prediction aims to infer the probability that users will click on advertisements, which is crucial for platforms to plan and optimize the expected revenues in online advertising. Substantial research efforts have been invested to develop a variety of CTR prediction models. However, there remains a lack of standardized benchmarks for validating and comparing the performance of CTR prediction models, especially in the situation with numerous models developed by researchers and practitioners. This research designs a unified architecture of CTR prediction benchmark (Bench-CTR) platform that offers flexible interfaces with datasets and components of a wide range of CTR prediction models. Moreover, we construct a comprehensive system of evaluation protocols encompassing real-world and synthetic datasets, a taxonomy of metrics, standardized procedures and experimental guidelines for calibrating the performance of CTR prediction models. Furthermore, we implement the proposed benchmark platform and conduct a comparative study to evaluate a wide range of state-of-the-art models from traditional multivariate statistical to modern large language model (LLM)-based approaches on three public datasets and two synthetic datasets. Experimental results reveal that, (1) high-order models largely outperform low-order models, though such advantage varies in terms of metrics and on different datasets; (2) LLM-based models demonstrate a remarkable data efficiency, i.e., achieving the comparable performance to other models while using only 2% of the training data; (3) the performance of CTR prediction models has achieved significant improvements from 2015 to 2016, then reached a stage with slow progress, which is consistent across various datasets. This benchmark is expected to facilitate model development and evaluation and enhance practitioners' understanding of the underlying mechanisms of models in the area of CTR prediction. Code is available at https://github.com/NuriaNinja/Bench-CTR.

# 1. Introduction

Online advertising revenue in the United States grew by 60.9% in 2023 compared to 2020, increasing from $139.8 billion to $225 billion (Statista, 2024). The online advertising market size is estimated at $257.97 billion in 2024 and is expected to reach $431.76 billion by 2029, growing at a compound annual growth rate (CAGR) of 10.85% during the forecast period from 2024 to 2029 (Mordor Intelligence, 2024). Click-through rate (CTR) prediction is essential for advertisers to optimize their decisions and maximize the expected revenues from online advertising campaigns.

In the literature, researchers have developed plenty of CTR prediction models in the last decade, including multivariate statistical models (e.g., Chang et al., 2010; Kumar et al., 2015), tree-based models (e.g., Friedman, 2001; Chen & Guestrin, 2016), factorization machines (FMs)-based models (e.g., Rendle, 2010; Juan et al., 2016; Pan et al., 2018; Sun et al., 2021), deep learning models (e.g., Qu et al., 2016; Yang et al., 2020; Niu & Hou, 2020; Chen et al., 2022b; Zhai et al., 2023), ensemble models (e.g., Cheng et al., 2016; Lian et al., 2018; Cheng et al., 2020; Wang et al., 2023b; Liu et al., 2024), and large language model (LLM)-based models (e.g., Yang et al., 2023a; Fu et al., 2023; Geng et al., 2024). However, the impressive number of models and a variety of evaluation procedures and metrics have made rigorous model evaluation quite challenging. In particular, studies employing non-uniform and incompatible procedures lead to non-reproducible, inconsistent or even opposing results (Zhu et al., 2021; Yang & Zhai, 2022), and in turn limit the credibility of CTR prediction models. The inherent sparsity and inadequate field documentation in public datasets hinder their effectiveness for research. Moreover, the disparate selection of evaluation metrics among studies limits the comparability of results. Thus, there lacks a standardized benchmark for validating and comparing the performance of CTR prediction models, along with synthetic data generation and metrics selection strategies, which motivates us to develop a comprehensive benchmark for CTR prediction in online advertising. To the best of our knowledge, this is one of the first research efforts in this direction.

This research has three objectives. First, we aim to design a unified architecture of CTR prediction benchmark (Bench-CTR) platform that offers flexible interfaces with datasets and components of a wide range of CTR prediction models. Second, we intend to construct a comprehensive system of evaluation protocols encompassing real-world and synthetic datasets, a taxonomy of metrics, standardized procedures and experimental guidelines for calibrating the performance of CTR prediction models. Third, our goal is to implement Bench-CTR and conduct a comparative study to evaluate a wide range of state-of-the-art models from traditional multivariate statistical to modern LLM-based approaches.

This research complements recent studies on evaluation frameworks (Zhu et al., 2021) and literature reviews (Yang & Zhai, 2022) in this area. Specifically, Zhu et al. (2021) focused on an evaluation environment with two public datasets (i.e., Criteo and Avazu), two metrics (i.e., AUC and Logloss) and basic toolkits (i.e., data splitting and preprocessing, hyper-parameters tuning and interfaces with CTR prediction models); and Yang and Zhai (2022) presented a state-of-the-art literature review on CTR prediction research focusing on modeling frameworks and techniques.

In this research, we propose a benchmark platform for CTR prediction (Bench-CTR) to facilitate fair and consistent comparisons among various models in the context of online advertising. First of all, we present a unified architecture of CTR prediction benchmark with essential interfaces with datasets and modeling components. Moreover, we provide a comprehensive system of evaluation protocols including synthetic data generation and real-world datasets publically available and widely used in the literature, a multi-level taxonomy of evaluation metrics, evaluation procedures and experimental guidelines for performance assessment of CTR prediction models. Furthermore, we provide implementation details of the proposed benchmark platform and synthetic data generation algorithms, and conduct a comparative study to validate 15 CTR prediction models in terms of 11 metrics on three public and two synthetic datasets. Results reveal that, (1) high-order models largely outperform low-order models, though such advantage varies in terms of metrics and on different datasets; (2) LLM-based models demonstrate a remarkable data efficiency, i.e., achieving comparable performance to other models while using only 2% of the training data; (3) the performance of CTR prediction models has achieved significant improvements from 2015 to 2016, then reached a stage with slow progress in the past decade, which is consistent across various datasets.

The contributions of this research can be summarized as follows. First, we propose a unified and extensible benchmark (Bench-CTR) platform that integrates flexible interfaces with datasets and state-of-the-art models for CTR prediction in the context of online advertising. Second, we establish a comprehensive system of evaluation protocols that encompasses a collection of public and synthetic datasets, a taxonomy of metrics, standardized procedures and guidelines. Third, we implement and employ Bench-CTR to perform a comparative study validating a wide range of CTR prediction models.

The remainder of this paper is organized as follows. Section 2 reviews related work on benchmarking studies. Section 3 presents a unified architecture of Bench-CTR and related interfaces and components. Section 4 gives the CTR prediction modeling process. Section 5 describes a comprehensive system of evaluation protocols. Section 6 provides implementation details of Bench-CTR and reports a comparative study conducted on Bench-CTR. In Section 7, we discuss main challenges and future directions in this area, and conclude the paper in Section 8.

## 2. Related work

Benchmarks serve as standard tools for evaluating and comparing competing systems or components in the scientific community. Benchmarking studies can be categorized into three main groups, including evaluation frameworks, comparative studies, and benchmarking datasets. For evaluation frameworks, MT-bench (multi-turn benchmark) (Zheng et al., 2023) is designed to evaluate large language models' multi-turn conversation and instruction-following capabilities, RecBole (Zhao et al., 2021) and Elliot (Anelli et al., 2021) focus on assessing the performance recommendation models. In the branch of comparative studies, researchers have conducted evaluations, including performance comparisons between predictive models and ChatGPT (Caruccio et al., 2024), as well as assessments of fine-tuned pretrained language models (Doimo et al., 2024). In the field of benchmarking datasets, CIFAR-10 (Canadian institute for advanced research) (Krizhevsky & Hinton, 2009) and ImageNet (Deng et al., 2009) in computer vision, ARC (AI2 reasoning challenge) (Clark et al., 2018), HellaSwag (harder endings, longer contexts, and low-shot activities for situations with adversarial generations) (Zellers et al., 2019), GLUE (general language understanding evaluation) and SuperGLUE (Wang et al., 2019a; 2019b) in natural languages provide standardized data for model evaluation.

Although abundant benchmarking studies have been reported in various fields as discussed above, to the best of our knowledge, there is little research on developing benchmarking and comparative studies in the area of CTR prediction, with the exception of Zhu et al. (2021). Our work differs from Zhu et al. (2021) in the following key aspects. This research proposes a unified architecture of CTR prediction benchmark (Bench-CTR) platform, which offers flexible interfaces with datasets and components of a wide range of CTR prediction models from traditional multivariate statistical to modern large language model (LLM)-based approaches, while Zhu et al.

(2021) narrowly focused on a simplified evaluation environment with two public datasets (i.e., Criteo and Avazu), two metrics (i.e., AUC and Logloss) and basic toolkits (i.e., data splitting and preprocessing, hyper-parameters tuning and interfaces with CTR prediction models). Moreover, we provide a comprehensive system of evaluation protocols, encompassing two synthetic data generation algorithms and three public datasets (i.e., Criteo, Avazu, and AntM$^2$C), a multi-level taxonomy of twenty evaluation metrics used in the literature on CTR prediction, and the evaluation procedure including dataset selection, validation methods, hyper-parameter settings, metric selection, model comparison, sensitivity analysis and ablation studies.

Besides, this research fundamentally differs from Yang and Zhai (2022) in the following key aspects. We propose a unified architecture of CTR prediction benchmark (Bench-CTR) platform embodying standardized evaluation protocols and flexible interfaces with datasets and modeling components supporting experimental studies and evaluation of CTR prediction models, while Yang and Zhai (2022) presented a state-of-the-art literature review on CTR prediction research focusing on modeling frameworks and techniques. Moreover, we conduct a comparative study covering state-of-the-art models across a wide range from traditional multivariate statistical models, tree-based models, factorization machines (FMs)-based models, deep neural network (DNN)-based models, convolutional neural network (CNN)-based models, recurrent neural network (RNN)-based models, graph neural network (GNN)-based models, transformer-based models, ensemble models, to modern LLM-based models.

## 3. The Bench-CTR architecture

The CTR prediction problem in online advertising pertains to estimating the likelihood that a user will click on an advertisement. The mathematical definition of the CTR prediction problem is given as $p(u, a) = f\big(\varphi(u, a; \theta)\big)$, where $p(u, a)$ is the probability that a user will click on an advertisement, $u$ is user, $a$ is an advertisement, $\varphi(u, a; \theta)$ is a function processing the information of user and advertisement, $\theta$ represents the parameters of this function, and $f$ is a function that maps the output of $\varphi(u, a; \theta)$ to a probability value which is between 0 and 1.

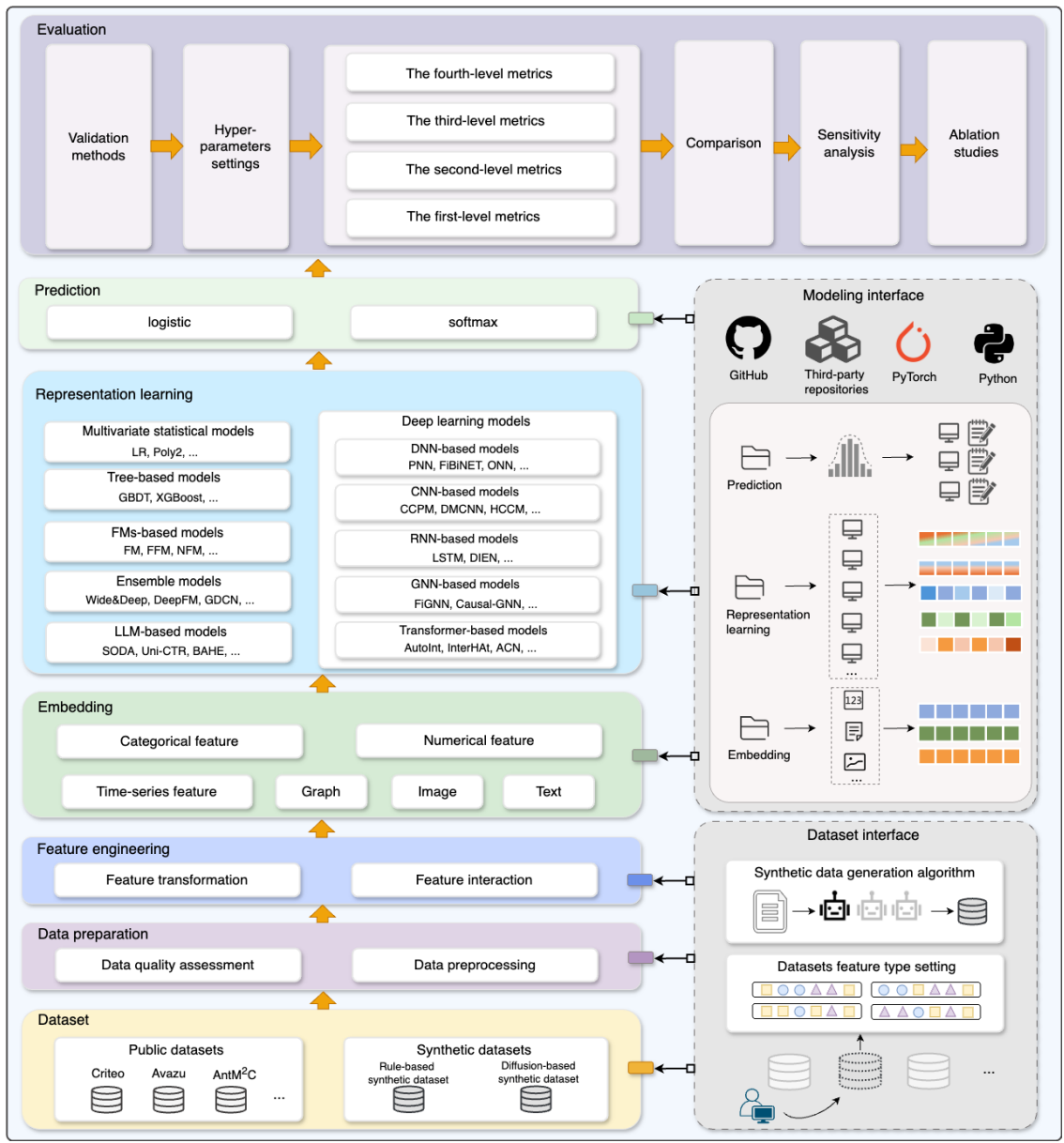

Figure 1. The architecture of Bench-CTR platform

In this research, we propose a unified architecture of CTR prediction benchmark (Bench-CTR) platform, as shown in Figure 1. Specifically, the architecture of Bench-CTR platform consists of three major components: datasets handler, CTR modeling suites, and evaluation system. The datasets handler provides synthetic data generation algorithms to produce synthetic datasets and an interface with public datasets (e.g., Criteo and Avazu) widely used in the extant literature. Public datasets ensure the comparability of experimental results reported in prior studies, while synthetic datasets offer the opportunities of controlling experimental processes. The dataset interface also incorporates data preparation and feature engineering functions. For the CTR modeling suites, Bench-CTR provides a set of modeling interfaces supporting various embedding, representation learning and the click probability prediction methods, by encapsulating source codes offered in prior studies and publically available third-party repositories. In other words, for each CTR prediction model, we decompose its modeling structure by modularizing key components (e.g., embedding and representation learning layers), and design an interface for each type of modules in Bench-CTR. The evaluation system contains a comprehensive protocol including validation methods, hyper-parameter settings, multi-level metrics, evaluation procedure and guidelines for performance comparison, sensitivity analysis and ablation studies.

Essentially, Bench-CTR consolidates the CTR prediction modeling process and a comprehensive system of evaluation protocols. The former is detailed in Section 4, and the latter is presented in Section 5.

## 4. The CTR prediction modeling process

Generally, the CTR prediction modeling process consists of five components, including data preparation, feature engineering, embedding, representation learning, and prediction, as shown in the bottom left of Figure 1.

### 4.1 Data preparation

Data preparation is a process that involves data quality assessment and preprocessing for consistency checking and data cleaning to ensure data quality for CTR prediction.

**Data quality assessment** provides information about various aspects of the dataset, such as data reliability, completeness and distribution, which can be conducted through exploratory data analysis (EDA) techniques, including box plots and correlation matrices.

**Data preprocessing** encompasses a series of operations, including duplicate sample removal, missing values imputation, column selection, data sampling, imbalance handling of label distributions (Zhang et al., 2020), and replacing unknown or out-of-vocabulary values with standardized tokens (e.g., <unknown> and OOV) (Guo et al., 2022a; Li et al., 2024b; Sang et al., 2024).

### 4.2 Feature engineering

Feature engineering primarily includes transforming existing features into new ones and designing feature interaction mechanisms in the extant research on CTR prediction.

**(1) Feature transformation**

• Discretization: numerical and categorical features can be transformed into discrete values by using equal-width and equal-frequency binning methods to simplify features and accelerate training process.

• Scaling and normalization: feature values can be rescaled to a specified range through min-max scaling (Wang et al., 2022a) and log transformation (Wang et al., 2017; Song et al., 2019) to reduce the impact of outliers and improve the model convergence.

**(2) Feature interaction**

• The first-order interaction: an individual feature independently influences the CTR prediction, as exemplified in logistic regression (LR) (Kumar et al., 2015).

• The second-order interaction: the effect of one feature on the CTR prediction depends on another one, as embodied in factorization machines (FMs) (Rendle, 2010).

• High-order interaction: the effect of one feature on the CTR prediction depends on two or more other ones, as described in deep learning models (e.g., Li et al., 2021a; Chen et al., 2022b; Zhai et al., 2023).

## 4.3 Embedding

Embedding transforms high-dimensional sparse inputs into low-dimensional dense vectors (Guo et al., 2017; Song et al., 2019), through the following techniques.

**(1) Categorical feature embedding** uses one-hot or multi-hot encoding to handle categorical features (Zhou et al., 2018).

**(2) Numerical feature embedding** uses a linear layer or an embedding table to encode scalar values.

**(3) Time-series feature embedding** uses transformation-based methods (e.g., Fourier transform and wavelet transform) or model-based methods (e.g., autoregressive model and hidden Markov model) to encode temporal features for the purpose of effectively capturing trends and patterns over time.

**(4) Text embedding** uses Doc2vec (Xia et al., 2019) or language models (Wang et al., 2022a) to encode text features.

**(5) Image embedding** uses method such as ResNet50 (Xia et al., 2019) and VGG16 (Li et

al., 2020b) to transform images into vectors that capture visual information.

**(6) Graph embedding** converts graph nodes or edges into vectors capturing structural information, by using method such as DeepWalk and node2vec. In CTR prediction, users and advertisements can be represented as nodes in a bipartite graph where edges describe the interactions between users and advertisements (Jin et al., 2025; Zhai et al., 2025).

## 4.4 Representation learning

Representation learning is referred to learning data representations during the model training process to make underlying explanatory factors more explicit and useful. In the literature, numerous models have been proposed to learn and utilize representations of users, advertisements and interactions effectively to improve CTR prediction. State-of-the-art CTR prediction models can be classified into six categories, including multivariate statistical models, tree-based models, FMs-based models, deep learning models, ensemble models, and LLM-based models, as summarized in Table 1. For details about CTR prediction modeling frameworks, refer to the literature by Yang & Zhai (2022). In Table 1, deep learning models include DNN-based, CNN-based, RNN-based, GNN-based and transformer-based models. While not exhaustive, Table 1 illustrates major categories of representative models for CTR prediction in the context of online advertising.

**(1) Multivariate statistical models**

• Logistic regression (LR) (Kumar et al., 2015) is a fundamental linear model widely used for CTR prediction, due to its simplicity and computational efficiency.

• Degree-2 polynomial model (Poly2) (Chang et al., 2010) expands linear models to capture non-linear relationships and pairwise interactions between features by assigning an independent parameter to each feature pair.

**(2) Tree-based models**

• Gradient boosting decision tree (GBDT) (Friedman, 2001) iteratively constructs an additive model, where each decision tree is trained to fit the negative gradients of the loss function of previously integrated trees.

• eXtreme gradient boosting (XGBoost) (Chen & Guestrin, 2016) incorporates a sparsity-aware algorithm for handling sparse data and a weighted quantile sketch algorithm for approximate tree learning, in order to prevent overfitting and improve the model performance.

**(3) FMs-based models**

- Factorization machines (FMs) (Rendle, 2010) efficiently capture pairwise feature interactions by factorizing associated parameters and are capable of generalizing to unseen feature combinations.

- Field-aware factorization machines (FFMs) (Juan et al., 2016) are a variant of FMs, considering field information to enhance the model performance in CTR prediction.

- Neural factorization machine (NFM) (He & Chua, 2017) integrates the second-order feature interactions modeling of FMs with the non-linearity of neural networks to capture high-order interactions.

- Attentional factorization machine (AFM) (Xiao et al., 2017) employs a neural attention network to learn the contributions of interactions through an attention-based pooling mechanism.

- Field-weighted factorization machine (FwFM) (Pan et al., 2018) captures feature interactions between features with fewer parameters compared to FFMs.

- High-order attentive factorization machine (HoAFM) (Tao et al., 2020) explicitly encodes high-order feature interactions into representations, and employs a bit-wise attention mechanism to determine the importance of feature interactions.

- Field-matrixed factorization machine (FmFM) (Sun et al., 2021) describes pairwise interactions with matrices, and improves the inference efficiency through caching intermediate vectors.

**(4) Deep learning models**

**DNN-based models** leverage multiple hidden layers of neural networks with non-linear transformations to learn high-order feature interactions implicitly.

- Product-based neural network (PNN) (Qu et al., 2016) consists of three components, namely an embedding layer to learn distributed representations, a product layer to capture the second-order interactions between features and a set of fully connected layers to represent high-order interactions.

- Feature importance and bilinear feature interaction network (FiBiNET) (Huang et al., 2019) employs a squeeze-excitation network (SENET) mechanism to dynamically learn the importance of features, and leverages a bilinear function to learn feature interactions.

- Operation-aware neural networks (ONN) (Yang et al., 2020) learn various representations

for different operations (e.g., product and convolution) to improve the prediction through the operation-aware embedding, a feature extraction layer and multiple nonlinear layers.

**CNN-based models** utilize convolutional and pooling operations to extract local patterns and capture interactions with neighboring features.

• Convolutional click prediction model (CCPM) (Liu et al., 2015) uses convolutional neural networks to extract local-global features from an individual and sequential advertising impressions.

• Density matrix-based convolutional neural network (DMCNN) (Niu & Hou, 2020) combines density matrices and convolutional neural networks to capture high-order feature interactions.

• Hybrid CNN based attention with category prior module (HCCM) (Chen et al., 2022b) combines fixed-CNN and trainable-CNN components to model users' behaviors in CTR prediction.

**RNN-based models** employ recurrent neural networks to learn dynamic representations and temporal patterns from users' sequential behaviors.

• Long short-term memory (LSTM) (Hochreiter & Schmidhuber, 1997) is designed to address the limitation of recurrent neural networks in handling long-term dependencies by designing a gating mechanism to control information flow, which consists of input gate, forget gate, output gate and cell state.

• Deep interest evolution network (DIEN) (Zhou et al., 2019) encompasses an interest extractor layer that utilizes an auxiliary loss function to capture temporal interests and an interest evolving layer that employs gated recurrent unit (GRU) with an attentional update gate (AUGRU) to capture interest evolving processes.

**GNN-based models** capture feature interactions between nodes in a graph structure and learn representation through message aggregation and propagation.

• Feature interaction graph neural networks (FiGNN) (Li et al., 2019) utilize a graph structure to represent multi-field features, where a node denotes a feature field and an edge denotes a feature interaction.

• Causal-GNN (Zhai et al., 2023) leverages causal inferences to capture high-order interactions in a feature graph and combines representations from user-user and ad-ad graphs to enhance CTR prediction.

**Transformer-based models** employ an attention mechanism to capture long-range dependencies and facilitate parallel processing of sequential data.

• Automatic feature interaction learning (AutoInt) (Song et al., 2019) uses a self-attention mechanism to explicitly learn high-order feature interactions.

• Interpretable CTR prediction model with hierarchical attention (InterHAt) (Li et al., 2020c) employs a transformer with the multi-head self-attention to learn polysemy of feature interactions and a hierarchical attention structure to learn the importance of different orders of features.

• Attentive capsule network (ACN) (Li et al., 2021a) uses transformers to capture feature interactions and capsule networks to extract users' interests.

**(5) Ensemble models**

• Wide&Deep (Cheng et al., 2016) integrates wide linear models with deep neural networks to simultaneously achieve memorization and generalization.

• Deep&Cross network (DCN) (Wang et al., 2017) integrates capabilities of deep neural networks with a cross network architecture to learn bounded-degree feature interactions.

• Factorization-machine based neural network (DeepFM) (Guo et al., 2017) employs a hybrid architecture integrating deep neural networks and factorization machines to capture low-order and high-order feature interactions.

• eXtreme deep factorization machine (xDeepFM) (Lian et al., 2018) integrates compressed interaction network (CIN) with deep neural networks in a unified framework to capture feature interactions at the vector-wise level.

• Ensemble trees and cascading forests (ETCF) (Qiu et al., 2018) integrates GBDT with multi-grained cascade forest (gcForest), where features are transformed with GBDT and then fed to gcForest to improve the model performance.

• Deep learning framework distilled by GBDT (DeepGBM) (Ke et al., 2019) integrates deep neural networks with GBDT using specialized components to process categorical and numerical features.

• Adaptive factorization network (AFN) (Cheng et al., 2020) incorporates a log transformation with neural networks to learns arbitrary-order cross features and related weights adaptively.

• Feature selection and interaction aggregation layers on top of two multi-layer perceptron

(MLP) module networks (FinalMLP) (Mao et al., 2023) enhances two-stream MLP models with a feature gating layer and a bilinear interaction aggregation layer to improve CTR prediction.

• Gated deep cross network (GDCN) (Wang et al., 2023b) integrates a gated cross network (GCN) with DNN, where DNN and GCN are used to capture implicit and explicit high-order interactions, respectively, through an information gate.

• RLCP[1] (Liu et al., 2024) uses a reinforcement learning (RL) algorithm to dynamically determine weights while combining results of multiple CTR prediction models.

**(6) LLM-based models**

• SODA [2](Yang et al., 2023a) integrates ChatGPT based on GPT-3.5 (Ouyang et al., 2022) with explainable artificial intelligence techniques to improve CTR prediction and optimize advertising strategies.

• Unified framework for multi-domain CTR prediction (Uni-CTR) (Fu et al., 2023) employs a LLM to capture domain commonalities by extracting layer-wise representation and domain-specific networks to capture domain characteristics.

• Behavior aggregated hierarchical encoding (BAHE) (Geng et al., 2024) extracts representations of atomic behaviors by utilizing encoding techniques in a LLM (e.g., Qwen-1_8B[3]) to improve the computational efficiency.

Table 1. Categories of representative CTR prediction models

| Categories | Representative models | References |
| --- | --- | --- |
| Multivariate statistical models | LR | Kumar et al. (2015) |
| | Poly2 | Chang et al. (2010) |
| Tree-based models | GBDT | Friedman (2001) |
| | XGBoost | Chen & Guestrin (2016) |
| FMs-based models | FMs | Rendle (2010) |
| | FFMs | Juan et al. (2016) |
| | NFM | He & Chua (2017) |
| | AFM | Xiao et al. (2017) |
| | FwFM | Pan et al. (2018) |
| | HoAFM | Tao et al. (2020) |

---

[1] It is referred to a reinforcement learning-based ensemble learning framework for CTR prediction.
[2] It is referred to a LLM-based advertising analysis framework for CTR prediction and advertising strategy optimization.
[3] https://huggingface.co/Qwen/Qwen-1_8B (accessed on July 1, 2025)

| | | | |
|---|---|---|---|
| Deep learning models | DNN-based models | FmFM | Sun et al. (2021) |
| | | PNN | Qu et al. (2016) |
| | | FiBiNET | Huang et al. (2019) |
| | | ONN | Yang et al. (2020) |
| | CNN-based models | CCPM | Liu et al. (2015) |
| | | DMCNN | Niu & Hou (2020) |
| | | HCCM | Chen et al. (2022b) |
| | RNN-based models | LSTM | Hochreiter & Schmidhuber (1997) |
| | | DIEN | Zhou et al. (2019) |
| | GNN-based models | FiGNN | Li et al. (2019) |
| | | Causal-GNN | Zhai et al. (2023) |
| | Transformer-based models | AutoInt | Song et al. (2019) |
| | | InterHAt | Li et al. (2020c) |
| | | ACN | Li et al. (2021a) |
| Ensemble models | | Wide&Deep | Cheng et al. (2016) |
| | | DCN | Wang et al. (2017) |
| | | DeepFM | Guo et al. (2017) |
| | | xDeepFM | Lian et al. (2018) |
| | | ETCF | Qiu et al. (2018) |
| | | DeepGBM | Ke et al. (2019) |
| | | AFN | Cheng et al. (2020) |
| | | FinalMLP | Mao et al. (2023) |
| | | GDCN | Wang et al. (2023b) |
| | | RLCP | Liu et al. (2024) |
| LLM-based models | | SODA | Yang et al. (2023a) |
| | | Uni-CTR | Fu et al. (2023) |
| | | BAHE | Geng et al. (2024) |

## 4.5 Prediction

The predicted probability needs to be constrained to the range $[0, 1]$, by using functions such as logistic and softmax. The logistic function is the most widely adopted by researchers (e.g., Huang et al., 2019; Li et al., 2020c; Wang et al., 2022b; Zhai et al., 2023; Wang et al., 2023c), which is defined as follows.

$$\sigma(x) = \frac{1}{1+e^{-x}}. \quad (1)$$

An alternative method is softmax (Liu et al., 2015), which is defined as follows.

$$\text{softmax}(x_i) = \frac{e^{x_i}}{\sum_{j=1}^{2} e^{x_j}}, \quad (2)$$

where $i$ indicates the class index, and $j$ is the summation index.

# 5. A comprehensive system of evaluation protocols for CTR prediction

In this section, we provide a comprehensive system of evaluation protocols for CTR prediction, as shown in the top of Figure 1, including public and synthetic datasets (Section 5.1), a multi-level taxonomy of metrics (Section 5.2), evaluation procedures and experimental guidelines (Section 5.3).

## 5.1 Datasets

In this section, we introduce public datasets widely used in the literature on CTR prediction and synthetic datasets generation methods, which both are supported in Bench-CTR. Public datasets are collected from actual activities and advertising logs where a large number of features may be sparse, which requires manual feature engineering by domain experts and in turn hampers the prediction. In contrast, synthetic datasets are created based on predefined rules or generative models, allowing researchers to conduct controllable experiments.

### 5.1.1 Public datasets

Public datasets are usually released by companies or institutions, which record user interactions with advertisements in real practices. In the following we introduce nine datasets that are most commonly adopted or recently released for CTR prediction evaluation.

**Criteo**[4]. This dataset consists of a training set spanning 7 days and a test set spanning 1 day, which provides a total of 45.8 million instances with 13 numerical fields and 26 categorical fields.

**Avazu**[5]. This dataset consists of a training set spanning 10 days and a test set spanning 1 day, which contains 40.4 million records across 24 fields, such as banner position, site ID, site category, device type, click, and others.

**AntM$^2$C**[6]. This multi-modal dataset was provided by the Alipay platform in 2023, which includes 110 million impression-click samples across 37 fields, such as user ID, item ID, item content, and others.

---

[4] https://www.kaggle.com/c/criteo-display-ad-challenge (accessed on July 1, 2025)
[5] https://www.kaggle.com/c/avazu-ctr-prediction/data (accessed on July 1, 2025)
[6] https://www.atecup.cn/dataSetDetailOpen/1 (accessed on July 1, 2025)

**KDD12**[7]. This collection of training instances was derived from session logs of Tencent's proprietary search engine (i.e., soso.com), including query data and user information. Each record contains 12 fields such as UserID, AdID, Query, Depth, Position, Impression, and Click.

**iPinYou**[8]. This dataset contains advertising logs from iPinYou's proprietary demand-side platform (DSP) contain records of ad bidding, impressions, clicks, and conversions derived from real-world advertising campaigns. Each bidding line contains a unique bid ID that identifies an opportunity of ad impression and can be used to join impression and click logs. The landing page URL tracks users' clicks after they interact with an ad.

**Avito**[9]. This dataset was provided by the 2015 Kaggle click-through rate prediction competition, which is featured eight data tables from Avito, Russia's largest classifieds and commodity trading platform. The dataset samples users' information and search activity on avito.ru. For each user, one target impression between May 12 and May 20 is selected randomly as the test instance. In this dataset, the testing data are the collection of these target impressions, the training data includes all users' activities during at least 16 consecutive days from April 25 until the target impression.

**Outbrain Click Prediction**[10]. This dataset contains users' activities across multiple publisher sites in the United States, which consists of over 2 billion records of page views.

**Ali-CCP**[11]. This dataset was collected from logs of the recommender system in Taobao. Every row in the logs represents an impression, and contains fields, such as user ID, gender, age, shop ID, and others. In this dataset, the training set and the testing set are split along the time sequence.

**Aliyun Taobao display advertising dataset**[12]. This dataset includes 26 million ad display and click logs collected from 1.14 million sampled users on Taobao during 8 days, containing ad information, user profiles, and user shopping behavior records.

### 5.1.2 Synthetic datasets

Bench-CTR is equipped with two synthetic data generation algorithms for CTR prediction. The former produces users' behaviors and click-through records based on predefined data distributions, such as predefined interaction patterns (Tian et al., 2023), and conditional probability rules that feature groups jointly affect CTR prediction with predefined probabilities (Li et al., 2020c; Cao et al., 2021). The latter creates synthetic data by leveraging generative models, such as the synthetic minority over-sampling technique (SMOTE), conditional tabular generative adversarial network (CTGAN), tabular variational autoencoder (TVAE), denoising diffusion probabilistic model (DDPM), latent diffusion model (Xu et al., 2019; Zhang et al., 2024b), and data distillation (Wang et al., 2023a). We will present details about the two synthetic data generation algorithms in Section 6.1.2.

## 5.2 A multi-level taxonomy of evaluation metrics

In this section, we introduce mathematical definitions for evaluation metrics for CTR prediction, and discuss their advantages and disadvantages, as well as formal relationships among them.

### 5.2.1 Definitions of metrics

Metrics are quantitative measures to calibrate the performance of CTR prediction models. According to their definitions, the metrics can be organized in a hierarchical framework with four levels.

**(1) The first-level metrics** are the fundamental measures, upon which high-level metrics are constructed.

• True positive (TP) is defined as the number of instances where the model correctly predicts a click, i.e., when both the predicted result and the actual label indicate a click.

• False positive (FP) is defined as the number of instances where the model incorrectly predicts a click, i.e., when the predicted result indicates a click, but the actual label indicates a non-click.

• True negative (TN) is defined as the number of instances where the model correctly predicts a non-click, i.e., when both the predicted result and the actual label indicate a non-click.

• False negative (FN) is defined as the number of instances where the model incorrectly predicts a non-click, i.e., when the predicted result indicates a non-click, but the actual label indicates a click.

• Predicted probability is defined as the likelihood value predicted by the model, i.e., the

probability that a user will click on an advertisement.

**(2) The second-level metrics** are constructed based on the first-level metrics.

• Precision is defined as the number of true positives divided by the number of positive predictions, i.e., the proportion of ads predicted to be clicked that are actually clicked by users, which is given as follows.

$$Precision = \frac{TP}{TP+FP}. \quad (3)$$

Precision is useful in scenarios where false positives (e.g., non-clicks predicted as clicks) have detrimental effects (e.g., overestimating clicks); however, it ignores false negatives (i.e., undetected clicks).

• Recall is defined as the number of true positives divided by the number of actual positive instances, i.e., the proportion of actual clicks correctly predicted by the model, which is given as follows.

$$Recall = \frac{TP}{TP+FN}. \quad (4)$$

Recall is useful in scenarios where false negatives (e.g., clicks predicted as non-clicks) have detrimental effects (e.g., underestimating clicks); however, it ignores false positives, which may result in a model with high recall while generating a number of incorrect predictions.

• Accuracy is defined as the proportion of all predictions that are correct, which is given as follows.

$$\text{Accuracy} = \frac{TP+TN}{TP+FP+TN+FN}. \quad (5)$$

Accuracy provides a balanced assessment of model performance by considering positive and negative predictions; however, it is not suitable for imbalanced datasets.

• Matthews correlation coefficient (MCC) takes TP, TN, FP, and FN into account, and its value ranges from -1 to 1, where 1 indicates a perfect prediction, 0 indicates a prediction no better than random guessing, and -1 indicates a completely inverse prediction, which is given as follows.

$$\text{MCC} = \frac{TP \times TN - FP \times FN}{\sqrt{(TP+FP) \times (TP+FN) \times (TN+FP) \times (TN+FN)}}. \quad (6)$$

MCC provides a more balanced evaluation of model performance by considering all components of the confusion matrix containing predicted and actual results for clicks and non-clicks; however, it is less intuitive than accuracy.

• Logloss is defined as the divergence between predicted probabilities and actual users'

behaviors, and its value ranges from 0 to infinity, where the higher value indicates the greater divergence, which is given as follows.

$$\text{Logloss} = -\frac{1}{N}\sum_{i=1}^{N}(y_i \log(p_i) + (1-y_i) \log(1-p_i)), \quad (7)$$

where $N$ is the number of instances, $y_i$ is the actual value (i.e., click and non-click) for the $i$-th instance, and $p_i$ is the predicted probability of the $i$-th instance.

Logloss measures model performance by heavily penalizing incorrect predictions with high confidence score; however, it is sensitive to outliers.

• Mean square error (MSE) quantifies the average squared difference between the predicted probabilities and the actual users' behaviors. The lower value of MSE indicates the better model performance. It is given as follows.

$$\text{MSE} = \frac{1}{N}\sum_{i=1}^{N}(y_i - p_i)^2. \quad (8)$$

MSE measures the model performance by heavily penalizing larger prediction errors; however, it is sensitive to outliers and is difficult to interpret in classification tasks.

• Click over predicted click (COPC) is defined as dividing the actual click rate by the predicted click rate to assess models' tendency of overestimating or underestimating clicks. The value of COPC less than 1 indicates that the model overestimates clicks. It is given as follows.

$$\text{COPC} = \frac{\sum_{i=1}^{N} y_i}{\sum_{i=1}^{N} p_i}. \quad (9)$$

COPC assesses model performance by predicting click volumes (Chen et al., 2022a); however, it does not account for the calibration performance, leading to an incomplete assessment.

• Kullback-Leibler divergence (KLD) is defined as the divergence between the predicted click probabilities and the actual probabilities. The lower KLD value indicates that the predicted probabilities are closer to the actual distribution. It is given as follows.

$$\text{KLD} = \sum_{i=1}^{N} y_i' \log\frac{y_i'}{p_i'}, \quad (10)$$

where $y_i'$ is a normalized actual value, $p_i'$ is a normalized predicted value, and $N$ is the total number of instances.

KLD measures the difference between predicted and actual click probability distributions; however, it is sensitive to outliers.

• Field-level calibration error measures the difference between predicted and observed CTR within specific field segments, including field-level expected calibration error (Field-ECE) and field-level relative calibration error (Field-RCE). Specifically, Field-ECE is defined as the weighted sum of the predicted average error in each segment, and Field-RCE is defined as the average error rate divided by the actual values, which are given as follows.

$$\text{Field-ECE} = \frac{1}{|\mathcal{D}|} \sum_{z=1}^{|Z|} \left| \sum_{i=1}^{|\mathcal{D}|} (y_i - p_i) \, \mathbf{1}_{[z_i = z]} \right|, \quad (11)$$

$$\text{Field-RCE} = \frac{1}{|\mathcal{D}|} \sum_{z=1}^{|Z|} N_z \frac{\left| \sum_{i=1}^{|\mathcal{D}|} (y_i - p_i) \mathbf{1}_{[z_i = z]} \right|}{\sum_{i=1}^{|\mathcal{D}|} (y_i + \epsilon) \mathbf{1}_{[z_i = z]}}, \quad (12)$$

where $|\mathcal{D}|$ is the total number of instances, input space is divided into $|Z|$ disjoint segments, $N_z$ is the number of instances in each segment, and $\epsilon$ is a small positive constant to prevent division by zero.

Field-level calibration error provides consistent and reliable assessments by evaluating calibration among specific fields (Guo et al., 2022b); however, it requires manual selection of fields and fail to account for unselected fields.

**(3) The third-level metrics** are constructed based on the second-level metrics.

• Area under ROC curve (AUC-ROC) is defined as the area under the receiver operating characteristic (ROC) curve by illustrating the trade-off between the true positive rate and the false positive rate across different thresholds, and its value ranges from 0 to 1, where the higher value indicates better model performance. Note that TPR is identical to recall, as defined in Equation (4). FPR is defined as follows.

$$FPR = \frac{FP}{FP + TN}. \quad (13)$$

AUC-ROC handles imbalanced datasets better than accuracy; however, it may be overly optimistic when the dataset is extremely imbalanced.

• Area under precision recall curve (AUC-PR) is defined as the area under the curve by illustrating the trade-off between precision and recall across different thresholds. The higher AUC-PR value indicates the model can effectively balance precision and recall.

AUC-PR is suitable for imbalanced datasets; however, it ignores true negatives.

• F1 score is defined as the harmonic mean of precision and recall, which is given as follows.

$$F1 = \frac{2 \times precision \times recall}{precision + recall}. \quad (14)$$

F1 score balances precision and recall, enabling the evaluation of model's ability to predict potential clicks while minimizing errors; however, it represents model performance at a specific threshold, which may not capture the overall model performance.

• Root mean square error (RMSE) is defined as the square root of the MSE. The lower RMSE value indicates better model performance. It is given as follows.

$$RMSE = \sqrt{\frac{1}{N}\sum_{i=1}^{N}(y_i - p_i)^2}. \quad (15)$$

RMSE heavily penalizes larger errors through squaring operation, which aligns with CTR prediction tasks where mispredictions impact advertising campaign performance; however, it is sensitive to outliers.

• Relative information gain (RIG) is defined as the relative improvement of model performance compared to a baseline model by comparing their Logloss values, ranging from 0 to 1. The higher RIG value indicates better model performance. It is given as follows.

$$RIG = 1 - \frac{Logloss_{model}}{Logloss_{baseline}}. \quad (16)$$

RIG provides a normalized measure of performance improvement over a baseline model; however, it depends on the selection of a baseline model, which can influence the results.

**(4) The fourth-level metric** is constructed based on the third-level metrics.

• Relative improvement (RelaImpr) is defined as the relative improvement of model performance compared to a baseline model by comparing their AUC-ROC values. The RelaImpr value greater than 0 indicates the model outperforms the baseline model. It is given as follows.

$$RelaImpr = \frac{AUC_{model} - 0.5}{AUC_{baseline} - 0.5} - 1. \quad (17)$$

RelaImpr measures the model performance relative to a baseline model; however, it requires manual selection of a baseline model, which affects the results.

## 5.2.2 Categories and relationships among metrics

Evaluation metrics introduced in Section 5.2.1 can be categorized into two groups, namely confusion matrix-based and probability-based metrics. Specifically, confusion matrix-based metrics include precision, recall, accuracy, MCC, AUC-ROC, AUC-PR, F1 score, and RelaImpr, which are constructed based on components of confusion matrix (i.e., TP, FP, TN, and FN); probability-based metrics include Logloss, field-level calibration error, KLD, COPC, MSE, RMSE, and RIG, which are constructed based on the predicted probability. Based on mathematical definitions of these metrics introduced in Section 5.2.1, we can draw relationships among them, as shown in Figure 2.

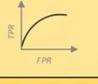

Figure 2. Relationships among evaluation metrics for CTR prediction

## 5.3 The evaluation procedure

The evaluation procedure includes dataset selection, validation, hyper-parameter settings, metrics selection, comparison, sensitivity analysis, and ablation studies, as shown in Figure 3.

### 5.3.1 Dataset selection

In CTR prediction, researchers need to select the most suitable datasets to fulfil their research objectives, considering the following criteria. First, the advertising formats (e.g., search ads and display ads) associated with the selected datasets should be consistent with the application scenario considered by CTR prediction models. However, it is worthwhile to note that a large portion of existing research fails to capture the specifics of advertising types in their models (Yang and Zhai, 2022) and thus ignores this principle of dataset selection. Second, some CTR prediction models take advantage of a set of specified features, for example, LLM-based models largely need textual features (Fu et al., 2023), and MLLM (multi-modal large language model)-based models rely on image or video inputs. Third, CTR prediction models based on semantics and knowledge graphs may need explicit definitions of features in datasets (Jin et al., 2025). Lastly, it is necessary to consider the popularity of datasets which may facilitate experimental comparisons and in turn convince other researchers and practitioners.

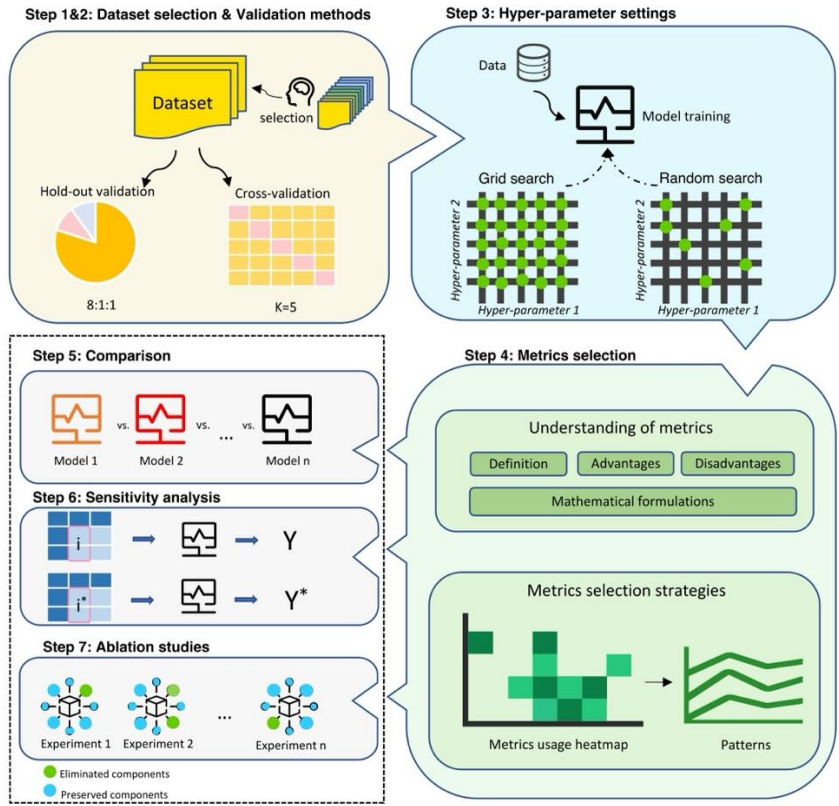

Figure 3. The evaluation procedure for CTR prediction

## 5.3.2 Validation methods

In order to obtain reliable performance of CTR prediction models, researchers employ validation methods to split a given dataset into statistically independent training, validation and testing sets. Specifically, training set enables models to learn patterns from data, validation set is employed for hyper-parameter tuning and overfitting detection, and testing set provides the final evaluation of model performance. Technically, validation methods are used to assess the model generalization on unseen data and prevent optimistic bias. In the literature on CTR prediction, there are two most widely used validation methods, including hold-out validation and cross-validation.

**Hold-out validation** splits a dataset into distinct subsets in two ways. One way is to randomly split the dataset in the ratio such as 8:1:1 or 7:2:1, representing the proportions for training, validation, and testing, respectively (Cheng et al., 2020; Li et al., 2024b; Sang et al., 2024). Another is chronological splitting, where the dataset is divided into multiple periods based on intervals (e.g., days, weeks) or the sequence of users' behaviors (Qin et al., 2020; Guo et al., 2022a; Wang

et al., 2022a).

**Cross-validation** splits the datasets into $k$ groups. Each group is iteratively used as the validation set and the other groups serve as the training set. The final model performance is calculated as the average results of $k$ runs, where $k$ is commonly set to 5 or 10 (Kumar et al., 2015).

### 5.3.3 Hyper-parameter settings

The performance of CTR prediction models is affected by settings of various hyper-parameters (e.g., learning rate and embedding size). Consequently, researchers conduct hyper-parameter optimization to identify the optimal setting of hyper-parameters from multiple parameter combinations, by using grid search (or random search).

**Grid search** exhaustively traverses hyper-parameter settings from a grid of specified hyper-parameter values.

**Random search** randomly samples hyper-parameter settings from predefined distributions over possible hyper-parameter values.

### 5.3.4 Metrics selection

In the literature, researchers have employed a variety of evaluation metrics to provide a quantitative measure of the model performance. However, it is not straightforward to select appropriate metrics from a large amount of options, as introduced in Section 5.2.1.

In the following, we summarize evaluation metrics used in the literature on CTR prediction in online advertising. Figure 4 presents a heatmap illustrating the frequency of metric usage in articles published from 2012 to 2024, where blue cells correspond to probability-based metrics and yellow cells correspond to confusion matrix-based metrics, with darker shades indicating higher usage frequencies. It is important to note that each percentage represents the proportion of articles that used a given metric out of those published in a specific year, and the percentages typically sum to equal or more than 100% in a specific year because individual articles may employ multiple metrics simultaneously. The original counts and percentages underlying Figure 4 are provided in Appendix A.2.

From Figure 4, we can observe the following phenomena in metrics selection: (1) AUC-ROC and Logloss are the most commonly used metrics for CTR prediction; specifically, a high usage rate of AUC-ROC has been maintained from 2012 to 2024, reaching a maximum rate of 100% in 2013 and 2023, and the usage rate of Logloss has increased from 25% in 2012 to 68% in 2023, indicating a growing acceptance of Logloss as evaluation metric by researchers; (2) the remaining confusion matrix-based metrics excluding AUC-ROC had low usage rates over years, so did the remaining probability-based metrics excluding Logloss. In summary, AUC-ROC and Logloss are two complementary and reliable metrics for CTR prediction tasks, which be explained with mathematical definitions and formal relationships of evaluation metrics mentioned in Section 5.2. Specifically, on one hand, AUC-ROC provides a balanced performance assessment by measuring the trade-off between true positive rate and false positive rate across all thresholds. In contrast, AUC-PR excludes true negatives and thereby neglects the model's ability to identify non-clicks; meanwhile, although accuracy and MCC consider all components of the confusion matrix, the former produces less reliable results on imbalanced datasets and the latter offers less intuitive interpretation, compared to AUC-ROC. On the other hand, Logloss directly evaluates probability estimates while penalizing incorrect predictions with high confidence.

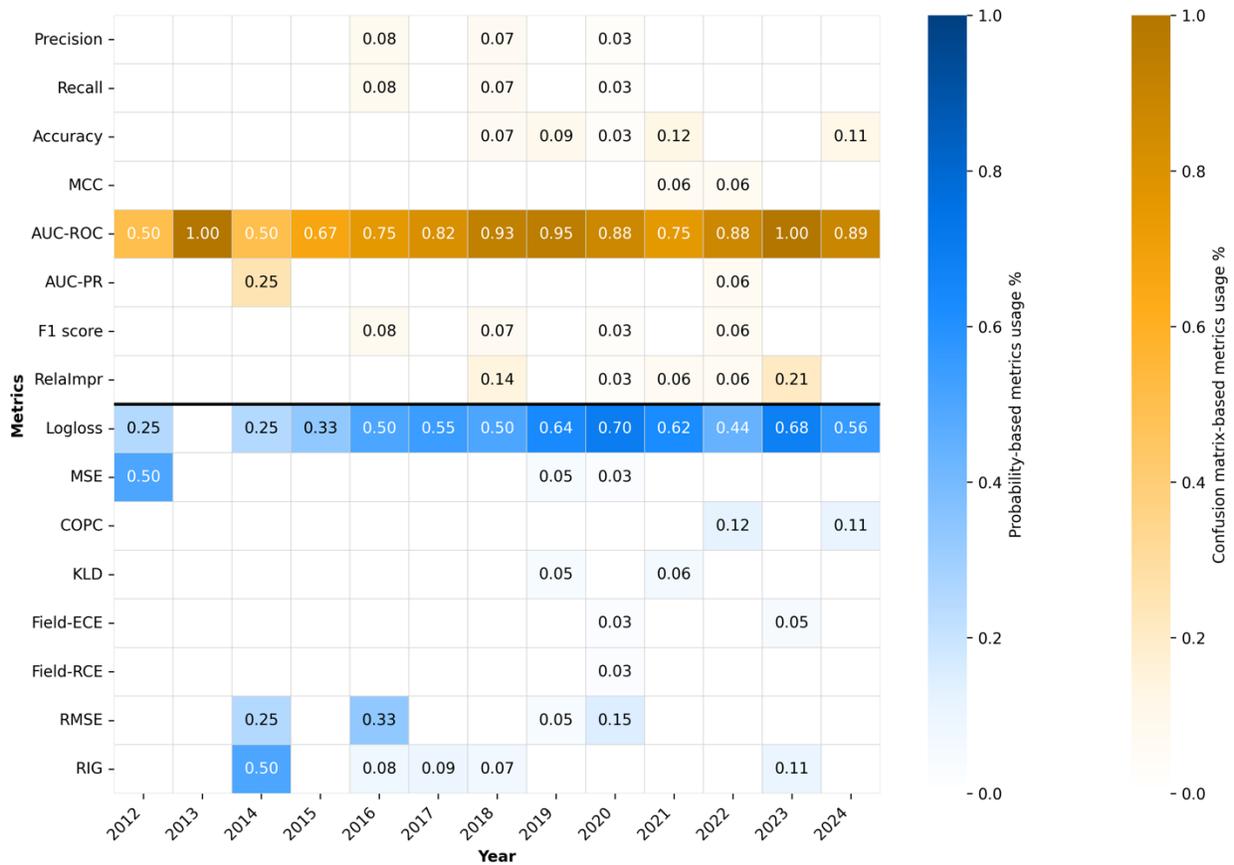

Figure 4. The heatmap of metrics used in the literature on CTR prediction (2012-2024)

### 5.3.5 Comparison

In the literature on CTR prediction, researchers have to conduct experimental comparisons to validate the effectiveness of their proposed models by comparing them with baseline models. However, there are a wide range of existing CTR prediction models. Hereby, we analyze baseline models utilized in the literature on CTR prediction in online, as shown in Figure 5. From Figure 5, we find that the evolution of baseline model selection can be divided into three periods. Specifically, in the first period (before 2016), LR and FMs have been commonly used as baseline models; in the second period (2017-2019), the set of baseline models have been expanded with Wide&Deep and DeepFM, and several additional models (e.g., AFM and NFM) employed by some studies to increase the model diversity; and in the third period (after 2020), the set of baseline models have been extended to include AutoInt, along with several additional models (e.g., PNN and xDeepFM). As for baseline model selection in CTR prediction, it is suggested to include baselines in the third period, namely LR, FMs, Wide&Deep, DeepFM, and AutoInt. Moreover, CTR prediction models recently developed in high-tier journals and conferences represent up-to-date advancements in this area, which should be included as obligatory baselines. Additionally, it is a wise option to prioritize models with publically available code to facilitate the reproducibility.

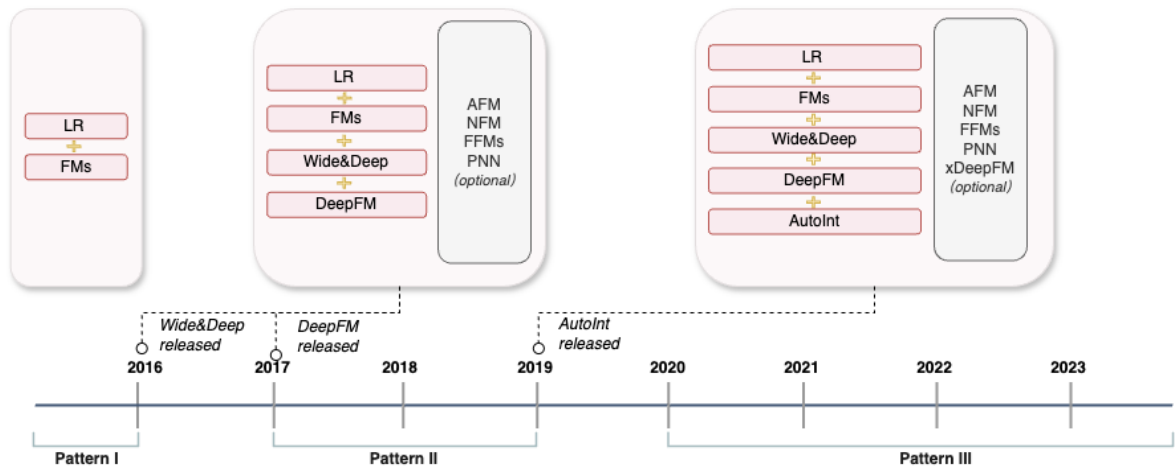

Figure 5. Evolution of baseline model selection over years

### 5.3.6 Sensitivity analysis

In CTR prediction research, sensitivity analysis has been conducted to explore how hyper-parameters affect the model performance. Table 2 shows the hyper-parameters usage in sensitivity analysis in the CTR prediction literature. Moreover, the hyper-parameters and their typical values are presented as follows. (1) Dropout is a regularization technique that randomly sets a fraction of input units to zero during model training. The dropout ratio is commonly selected from {0,0.1,0.2,...,0.9}. (2) Activation functions in neural networks introduce the non-linearity for learning complex interactions and patterns. In the literature on CTR prediction, there are three most widely used activation functions, including Sigmoid, ReLU, and Tanh. (3) The embedding size defines the dimensionality of embeddings, whose effect is typically explored in the range from 8 to 256. (4) The number of hidden layers represents the depth of a neural network, which commonly ranges from 1 to 7. (5) The number of hidden neurons refers to the number of units in each hidden layer, which is commonly set in the range from 10 to 2000. (6) The number of interacting layers is typically set in the range from 1 to 8, which are applied for modeling feature interactions.

Table 2. Hyper-parameters of CTR prediction models

| Hyper-parameters | References |
| --- | --- |
| The dropout ratio | Chen et al. (2016); Xiao et al. (2017); He & Chua (2017); Guo et al. (2017) |
| Activation functions | Qu et al. (2016); He & Chua (2017); Guo et al. (2017); Lian et al. (2018); Niu & Hou(2020) |
| The embedding size | Pan et al. (2018); Huang et al. (2019); Song et al. (2019); Tao et al. (2020); Zhai et al. (2023) |
| The number of hidden layers | Qu et al. (2016); Guo et al. (2017); Lian et al. (2018); Huang et al. (2019);Niu and Hou (2020); Cheng et al. (2020) |
| The number of hidden neurons | Guo et al. (2017); Lian et al. (2018); Huang et al. (2019); Niu & Hou(2020); Cheng et al. (2020) |
| The number of interacting layers | Blondel et al. (2016); Covington et al. (2016); Guo et al. (2017); Wang et al. (2017); Lian et al. (2018); Qu et al. (2018); Song et al. (2019); Cheng et al. (2020); Wang et al. (2023b) |

### 5.3.7 Ablation studies

In the literature on CTR prediction, researchers have to conduct ablation studies to evaluate the contribution of individual components to the full model performance in the following steps. The first step is to identify key components of the proposed model, e.g., attention mechanisms in the model of Li et al. (2020c) and adversarial learning in the model of Li et al. (2022). Subsequently, a set of variant models are coined by removing or modifying these components from the full model (Yang et al., 2023b; Yang et al., 2024; Zhang et al., 2024a). Finally, these variants are compared with the full model to evaluate how individual components contribute to the overall performance.

## 6. Bench-CTR implementation and the comparative study

We implement a benchmark platform for CTR prediction (Bench-CTR). In this section, we conduct a comparative study on Bench-CTR to validate 15 CTR prediction models in terms of 11 metrics on three public datasets and two synthetic datasets.

## 6.1 Implementation details of Bench-CTR

### 6.1.1 Platform implementation

We implement the Bench-CTR platform using Python 3.8.19 and PyTorch 2.4.1, following the architecture shown in Figure 1, with modular components for follow-up extensions. Specifically, the dataset interface loads public datasets and saves processed results into HDF5 files for efficient loading in subsequent steps, and also provides synthetic data generation algorithms to create datasets through configurable parameter settings; CTR prediction models have been encapsulated through the modeling interface in PyTorch; the training module supports mechanisms of learning rate decay, early stopping, and random seed management; the evaluation component supports 11 metrics with automatic logging of detailed experimental results, hyper-parameters and metric values.

### 6.1.2 Synthetic data generation algorithms

In the following we propose two synthetic data generation algorithms for CTR prediction in the Bench-CTR platform: one is rule-based and another is diffusion-based.

#### (1) The rule-based synthetic data generation

The rule-based synthetic data generation process consists of three steps, including feature generation, feature interaction and click label generation, as shown in Algorithm 1. In the first step, numerical and categorical features are generated for each instance $x$. Categorical features are generated following $x_f \sim Categorical(C_f, P_f)$, where $x_f$ is the value of categorical feature $f$ for instance $x$, $C_f$ represents the set of values for the categorical feature $f$, and $P_f$ represents the probability distribution. For numerical features, user's age is generated following a truncated normal distribution bounded between 0 and 100, and item's price is generated following a log-normal distribution bounded between 1 and 30,000. In the second step, we design a function to model different orders of feature interactions. Following Li et al. (2020c), feature interactions are defined at three levels, namely the first-, second- and third-orders. For each interaction order, in a predefined cluster of features, similarity scores between the generated features and predefined target features are calculated and multiplied together, and then weighted with the cluster's importance and summed to get the interaction score, i.e., $I_k(x) = \sum_{g \in G_k} w_g \prod_{f \in g} s(x_f, t_f)$, where $I_k(x)$ denotes the $k$-th order interaction score for instance $x$, $s(x_f, t_f)$ denotes the similarity score function, $t_f$ is the predefined target value, $w_g$ represents the feature interaction weight for the cluster $g$, and $G_k$ is the set of feature clusters for order $k$. In the third step, the final click probability is computed by combining the base probability with a weighted sum of interaction scores according to $p(y = 1|x) = \beta_0 + \sum_{k=1}^{K} \alpha_k I_k(x)$, where $p(y = 1|x)$ is the final click probability, $\beta_0$ is the base click-through rate set to 0.01 following Yang et al. (2017), $\alpha_k$ is the weight for different orders, and $K$ denotes the total order of feature interaction. The click label is sampled from a Bernoulli distribution according to $Y \sim Bernoulli(p(y = 1|x))$, where $Y$ is the click label.

---

**Algorithm 1** Rule-based synthetic dataset generation

---

**Input**: The total number of synthetic instances $N$, the base click ratio $\beta_0$.
**Output**: The synthetic dataset $D$ with features and click labels.
1:     $D := \emptyset$;
2:     **for** $i \leftarrow 1$ to N **do**
3:         **for** each $f \in categorical\_features$ **do**
4:            $x_{i,f} :=$ Sample from $Categorical(C_f, P_f)$;
5:         **end for**
6:         $age :=$ Sample from $TruncatedNormal([0,100])$;
7:         $price :=$ Sample from $LogNormal([1,30000])$;
8:         $Score := 0$;
9:         **for** $k \leftarrow 1$ to $K$ **do**
10:            $I_k := 0$;
11:            **for** each $g \in G_k$ **do**
12:               $cluster\_score := w_g \prod_{f \in g} s(x_{i,f}, t_f)$;
13:               $I_k := I_k + cluster\_score$;
14:            **end for**
15:            $Score := Score + \alpha_k I_k$;
16:         **end for**
17:         $p(y = 1|x) := \beta_0 + Score$;
18:         $Y :=$ Sample from $Bernoulli(p(y = 1|x))$;
19:         $D := D \cup \{(x_{i,1}, x_{i,2}, \ldots, x_{i,10}, Y)\}$;
20:     **end for**
21:     **return** $D$

---

**(2) The diffusion-based synthetic data generation**

The diffusion-based approach leverages a latent diffusion model for tabular data generation (Zhang et al., 2024b), integrating CTR prediction loss to effectively capture feature interactions and ~~users'~~ behavioral patterns relevant to CTR prediction tasks, as shown in Algorithm 2. The whole process consists of two phases, including training phase and generation phase.

During the training phase, the latent diffusion model learns the underlying data distribution and patterns related to CTR prediction through three steps. In the first step, a variational autoencoder (VAE) is employed to map the real CTR prediction dataset into a continuous latent space. Specifically, the VAE encoder network transforms features into latent representations through fully connected layers with ReLU (rectified linear unit) activation and dropout regularization. The reparameterization trick (e.g., reparameterization gradient estimator) is employed to sample the latent representations from a multivariate Gaussian distribution. The VAE decoder maps the latent representations back to the original feature space. In the second step, a diffusion model with a linear noise schedule is trained to learn the distribution of latent representations through forward and backward processes. In the third step, a CTR predictor is jointly trained to regulate the reconstruction process through CTR prediction loss.

During the generation phase, the synthetic dataset is produced through three steps. In the first step, random noise is sampled in the latent space and then the noise prediction network is applied to gradually transform the noise into realistic latent representations. In the second step, the realistic latent representations are decoded back to feature space through the VAE decoder. In the third step, the CTR predictor creates click labels for the synthetic features, ensuring the generated data maintain realistic behavioral patterns consistent with the distribution of the real-world dataset.

---

**Algorithm 2** Diffusion-based synthetic dataset generation

**Input**: The real-world dataset $X$, labels $Y$, the number of synthetic instances $M$.
**Output**: The synthetic dataset $D$ with features and click labels.

1:    // Initialization
2:    Initialize VAE, DiffusionModel and CTRPredictor
3:    Set loss weights: $\lambda\_recon$, $\lambda\_kl$, $\lambda\_diff$, $\lambda\_ctr$
4:    Set hyper-parameters: $T$, $\alpha\_step$, $threshold$
5:    // Training phase
6:    **for** epoch $\leftarrow$ 1 to $E$ **do**
7:       **for** each batch $(x, y) \in (X, Y)$ **do**
8:          // VAE forward pass
9:          $\mu, \log \sigma^2 := \text{encoder}(x)$;
10:        $z := \text{reparameterize}(\mu, \log \sigma^2)$;
11:        $x\_recon := \text{decoder}(z)$;
12:        // Diffusion forward pass
13:        $t := \text{sample\_timesteps(batch\_size)}$;
14:        $\varepsilon := \text{sample from } N(0, I)$;
15:        $z\_noisy := \text{add\_noise}(z, t, \varepsilon)$;
16:        $\varepsilon\_pred := \text{DiffusionModel}(z\_noisy, t)$;

```
17:            // CTR prediction
18:            $y\_pred := \text{CTRPredictor}(x\_recon)$;
19:            // Compute loss
20:            $L\_recon := \text{MSE}(x\_recon, x)$;
21:            $L\_kl := \text{KL\_divergence}(\mu, \log \sigma^2)$;
22:            $L\_diff := \text{MSE}(\varepsilon\_pred, \varepsilon)$;
23:            $L\_ctr := \text{BCE}(y\_pred, y)$;
24:            $L\_total := \lambda\_recon \times L\_recon + \lambda\_kl \times L\_kl + \lambda\_diff \times L\_diff + \lambda\_ctr \times L\_ctr$;
25:            // Parameters update
26:            Update all model parameters using $\nabla L\_total$
27:        end for
28:    end for
29:    // Generation phase
30:    $D := \emptyset$;
31:    for $i \leftarrow 1$ to $M$ do
32:        $z := \text{sample from } N(0, I)$;
33:        for step $\leftarrow 1$ to num_steps do
34:            $t := \max(1, \text{step} \times (T \,/\, \text{num\_steps}))$;
35:            $\varepsilon\_pred := \text{DiffusionModel}(z, t)$;
36:            $z := z - \alpha\_step \times \varepsilon\_pred$;
37:        end for
38:        // Generate features and labels
39:        $x\_synthetic := \text{decoder}(z)$;
40:        $p\_synthetic := \text{sigmoid}(\text{CTRPredictor}(x\_synthetic))$;
41:        $y\_synthetic := p\_synthetic > threshold$;
42:        $D := D \cup \{(x\_synthetic, y\_synthetic)\}$;
43:    end for
44:    return $D$
```

## 6.2 The comparative study

### 6.2.1 Models under comparison

In the comparative study, we validate the performance of 15 state-of-the-art models, including multivariate statistical models (i.e., LR), tree-based models (i.e., XGBoost), FM-based models (i.e., FM, AFM, NFM, and FmFM), DNN-based models (i.e., PNN), CNN-based models (i.e., CCPM), RNN-based models (i.e., LSTM), GNN-based models (FiGNN), transformer-based models (i.e., AutoInt), ensemble models (i.e., Wide&Deep, DeepFM, and GDCN) and LLM-based models (i.e., Uni-CTR).

### 6.2.2 Evaluation metrics

We measure the CTR prediction performance using 11 metrics, including confusion matrix-based metrics (i.e., AUC-ROC, AUC-PR, precision, recall, accuracy, MCC, and F1 score) and probability-base metrics (i.e., Logloss, MSE, RMSE, and COPC).

### 6.2.3 Datasets and data preparation

We use three public datasets (i.e., Criteo, Avazu, and AntM$^2$C) and two synthetic datasets in the comparison study. The statistics of the datasets are shown in Table 3.

Table 3. Summary statistics of Datasets

| Datasets | %Positive | #Training | #Validation | #Testing | #Fields |
|---|---|---|---|---|---|
| Criteo | 26% | 36,672K | 4,584K | 4,584K | 39 |
| Avazu | 17% | 32,343K | 4,043K | 4,043K | 24 |
| AntM$^2$C | 47% | 1,538K | 171K | 190K | 37 |
| Rule-based synthetic dataset | 25% | 25,694K | 3,212K | 3,212K | 10 |
| Diffusion-based synthetic data | 26% | 80K | 10K | 10K | 39 |

Note: K means thousand.

To ensure optimal model performance, we employ different data preprocessing methods for each dataset. For Criteo, missing values of numerical features and categorical features are assigned to zeros and null strings, respectively. Numerical features are transformed through a binning method, i.e., $f(x) = \lfloor (\ln(x))^2 \rfloor$ if $x$ exceeds a specified threshold, otherwise $f(x) = \lfloor x \rfloor$. Moreover, feature values with frequency less than 10 are mapped to an out-of-vocabulary (OOV) token, and others are mapped to unique tokens. For Avazu, missing values of categorical features are assigned to null strings. Temporal features are processed to obtain the hour of day, the day of week, and the weekend. For AntM$^2$C, we obtain instances of the advertising scenario by filtering the dataset with scene ID. Temporal features are processed in a similar way as for Avazu. Note that, for LLM-based models, missing values of numerical and categorical features are replaced with "unknown". Following Fu et al. (2023), we consolidate features into textual sequences with a defined prompt template. For the two synthetic datasets, we create 32,118K instances by using the rule-based data generation algorithm and 80K instances from the Criteo training data by using the diffusion-based data generation algorithm, and apply the same binning method as for Criteo.

### 6.2.4 Parameter settings

In the following experiments, we employ grid search to tune key hyper-parameters to ensure fair comparison across different models. Specifically, (1) we set the training process with a maximum of 99 epochs, incorporating the early stopping condition when there is no further improvement in the model's performance during 2 consecutive epochs on the validation dataset; (2) the learning rate is initialized at 0.001 using the Adam optimizer and searched with a manual reduction strategy; (3) for categorical features, the embedding dimension is set to 16 and the embedding regularization weight is set to 0.000001; (4) in order to make full use of GPU memory, we employ an adaptive batch size strategy initializing at 3,000 and progressively increasing in steps of 1,000 until encountering an out-of-memory error or reaching a maximum of 30,000; (5) the dropout rate is chosen from $\{0.0, 0.1, 0.2\}$, and the number of hidden layers is searched from $\{0,1,2,3\}$; (6) for the GNN-based models, the number of GNN layers is selected from $\{2,3,4\}$. Moreover, all models are evaluated twice using random seeds of 2019 and 2020, and the final metrics are calculated as the average results of the two runs. Additionally, the following experiments are performed on NVIDIA GeForce RTX 4090 GPUs.

## 6.3 Experimental results and analysis

### 6.3.1 Model performance on the public datasets

Experimental results on the three public datasets are presented in Tables 4, 5 and 6, respectively. It is worth noting that the LLM-based model (i.e., Uni-CTR) is not performed on Criteo and Avazu in that the two datasets do not contain textual features. Moreover, the models excluding Uni-CTR are trained on 1,538K instances from AntM$^2$C, while Uni-CTR is fine-tuned using low-rank adaptation (LoRA) (Hu et al., 2021) on 30K instances sampled from the same dataset due to limited computational resources. For fair comparison, we conduct zero-shot inference using Qwen2.5-1.5B[13] which serves as the language model backbone of Uni-CTR. From Tables 4, 5 and 6, several findings can be observed.

(1) On Criteo, DeepFM achieves the best results in the majority of evaluation metrics, followed by Wide&Deep; on Avazu, PNN outperforms other models in most metrics, followed by DeepFM; and on AntM$^2$C, GDCN largely achieves the best performance, followed by Wide&Deep. Notably, all of these models with the best and second-best performance are high-order feature interaction models.

(2) Correlations are observed between AUC-ROC and AUC-PR, between MCC and F1 score, as well as among Logloss, MSE, and RMSE, indicating that models performing well on one metric

---

[13]https://huggingface.co/Qwen/Qwen2.5-1.5B (accessed on July 1, 2025)

tend to perform well on the related ones.

(3) The LLM-based model achieves competitive performance compared to other models trained on much larger datasets, demonstrating a remarkable data efficiency. As shown in Figure 6, where the vertical dashed lines indicate the size of the training data, Uni-CTR achieves an AUC-ROC score of 0.8191 using only 2% of the training data used by other models (e.g., LR, FM, and GDCN). Furthermore, the poor performance of Qwen2.5-1.5B validates the significant improvement achieved by Uni-CTR.

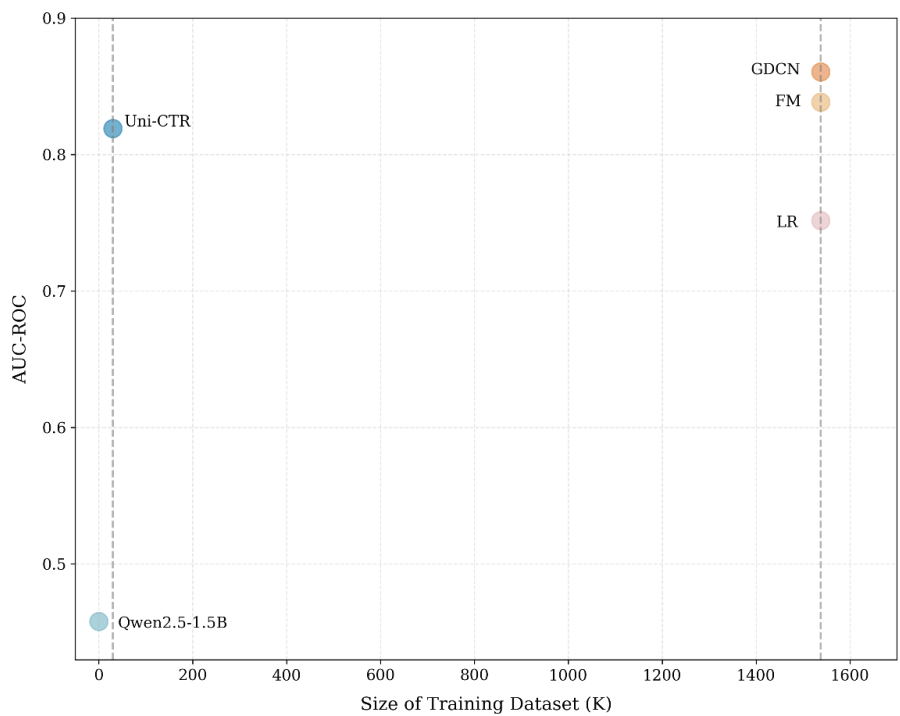

Figure 6: Performance of Uni-CTR and other models in terms of AUC-ROC

(4) It is apparent that COPC is not recommended as a standalone metric for performance evaluation, because Qwen2.5-1.5B achieves the best COPC score on AntM$^2$C, but perform worse than other models in terms of other metrics.

(5) Our experimental results reveal challenges in reproducing the relative performance between GDCN and DeepFM reported in the paper of Wang et al. (2023b). Specifically, the authors reported that AUC-ROC values of GDCN and DeepFM are 0.8161 and 0.8140, respectively, while in our study, AUC-ROC values of GDCN and DeepFM are 0.8114 and 0.8142, respectively; meanwhile, this phenomenon was also occurred to the metric of Logloss: the authors reported that Logloss values of GDCN and DeepFM are 0.4360 and 0.4378, respectively, while in our study, Logloss values of GDCN and DeepFM are 0.4404 and 0.4377, respectively. That is,

Wang et al. (2023b) reported that GDCN outperformed DeepFM, while our study got the opposite result, i.e., DeepFM outperformed GDCN. It is worthwhile to note that the result of our study is consistent with prior studies (e.g., Wang et al., 2025) and anecdotal evidence of un-replicability reported by researchers in the GDCN GitHub repository[14].

### 6.3.2 Model performance on the rule-based synthetic dataset

The experimental result on the rule-base synthetic dataset is presented in Table 7. Our findings are as follows. (1) The performance gaps among models are minimal, demonstrating that the synthetic dataset contains clear and learnable patterns that CTR prediction models of different architectures can understand and capture. This consistent model performance indicates that this algorithm generates stable and reliable datasets, making it suitable for researchers and practitioners to conduct controllable experiments under different configuration settings. (2) The result on the rule-based synthetic dataset reaffirms the correlation patterns observed in public datasets, particularly the relationships between AUC-ROC and AUC-PR, between MCC and F1 score, as well as among Logloss, MSE, and RMSE.

### 6.3.3 Model evaluation from synthetic to real-world datasets

In order to further evaluate the diffusion-based synthetic dataset, we adopt the TSTR (train on synthetic, test on real) (Esteban et al., 2017) procedure, i.e., training the CTR prediction models on the synthetic dataset and evaluating the model performance on the real-world dataset (i.e., Criteo). Experimental results of performance evaluation from synthetic to real-world datasets are presented in Table 8. We find that the AUC-ROC scores of models range from 0.63 to 0.65, demonstrating that the synthetic data can effectively capture underlying data distribution and click behavioral patterns from real-world datasets. While these metric scores are lower than those achieved by models trained on the full Criteo dataset (i.e., 36,672K instances), the performance gap is primarily attributed to the limited size of training data (i.e., 80K instances). It is worthwhile to note that the diffusion-based synthetic dataset is designed as complementary data to mitigate data sparsity and data privacy risks. The experimental results demonstrate the viability of the synthetic dataset for maintaining reasonable model performance and mitigating issues such as data sparsity and privacy risks.

---

[14] https://github.com/anonctr/GDCN/issues (accessed on July 1, 2025)

Table 4. Model performance comparison on Criteo

| Year | Feature interaction order | Model | Confusion matrix-based metrics | | | | | | Probability-based metrics | | | |
|------|------|------|------|------|------|------|------|------|------|------|------|------|
| | | | AUC-ROC | AUC-PR | Precision | Recall | Accuracy | MCC | F1 | Logloss | MSE | RMSE | \|1-CPOC\| |
| <2000 | 1 | LR | 0.7930 | 0.5848 | 0.6548 | 0.3554 | 0.7868 | 0.3670 | 0.4607 | 0.4571 | 0.1483 | 0.3851 | 0.0006 |
| <2000 | >2 | LSTM | 0.8048 | 0.6058 | 0.6524 | 0.4040 | 0.7921 | 0.3941 | 0.4990 | 0.4469 | 0.1446 | 0.3802 | 0.0091 |
| 2010 | 2 | FM | 0.8074 | 0.6087 | 0.6548 | 0.4059 | 0.7930 | 0.3968 | 0.5012 | 0.4447 | 0.1440 | 0.3794 | 0.0021 |
| 2015 | >2 | CCPM | 0.8101 | 0.6145 | 0.6628 | 0.4040 | 0.7946 | 0.4007 | 0.5020 | 0.4417 | 0.1429 | 0.3780 | 0.0019 |
| 2016 | >2 | XGBoost | 0.7714 | 0.5557 | 0.6530 | 0.2908 | 0.7787 | 0.3262 | 0.4024 | 0.4739 | 0.1543 | 0.3928 | **0.0003** |
| 2016 | >2 | PNN | 0.8134 | 0.6199 | 0.6632 | 0.4151 | 0.7961 | 0.4074 | 0.5106 | 0.4388 | 0.1419 | 0.3767 | 0.0055 |
| 2016 | >2 | Wide&Deep | <u>0.8141</u> | <u>0.6215</u> | 0.6629 | **0.4188** | <u>0.7965</u> | <u>0.4094</u> | **0.5133** | <u>0.4379</u> | <u>0.1416</u> | <u>0.3763</u> | 0.0138 |
| 2017 | 2 | AFM | 0.8046 | 0.6050 | 0.6595 | 0.3887 | 0.7919 | 0.3896 | 0.4891 | 0.4467 | 0.1447 | 0.3804 | <u>0.0004</u> |
| 2017 | >2 | NFM | 0.8089 | 0.6124 | 0.6586 | 0.4061 | 0.7939 | 0.3992 | 0.5024 | 0.4429 | 0.1433 | 0.3786 | 0.0039 |
| 2017 | >2 | DeepFM | **0.8142** | **0.6218** | 0.6637 | <u>0.4184</u> | **0.7967** | **0.4097** | <u>0.5132</u> | **0.4377** | **0.1415** | **0.3762** | 0.0129 |
| 2019 | >2 | FiGNN | 0.8137 | 0.6207 | **0.6642** | 0.4152 | 0.7964 | 0.4081 | 0.5110 | 0.4384 | 0.1417 | 0.3765 | 0.0030 |
| 2019 | >2 | AutoInt | 0.8116 | 0.6173 | 0.6613 | 0.4126 | 0.7953 | 0.4047 | 0.5081 | 0.4403 | 0.1424 | 0.3774 | 0.0057 |
| 2021 | 2 | FmFM | 0.8107 | 0.6152 | 0.6626 | 0.4060 | 0.7948 | 0.4017 | 0.5035 | 0.4412 | 0.1427 | 0.3778 | 0.0035 |
| 2023 | >2 | GDCN | 0.8114 | 0.6172 | <u>0.6640</u> | 0.4079 | 0.7954 | 0.4037 | 0.5053 | 0.4404 | 0.1424 | 0.3774 | 0.0142 |

Note: The best results across each metric column are **bolded**, the second-best results are <u>underlined</u>.

Table 5. Model performance comparison on Avazu

| Year | Feature interaction order | Model | Confusion matrix-based metrics | | | | | | | Probability-based metrics | | | |
|---|---|---|---|---|---|---|---|---|---|---|---|---|---|
| | | | AUC-ROC | AUC-PR | Precision | Recall | Accuracy | MCC | F1 | Logloss | MSE | RMSE | \|1-COPC\| |
| <2000 | 1 | LR | 0.7569 | 0.3869 | 0.6115 | 0.0828 | 0.8353 | 0.1804 | 0.1458 | 0.3927 | 0.1233 | 0.3511 | 0.0023 |
| <2000 | >2 | LSTM | 0.7694 | 0.3995 | 0.5836 | 0.1099 | 0.8355 | 0.2003 | 0.1849 | 0.3870 | 0.1217 | 0.3489 | 0.0117 |
| 2010 | 2 | FM | 0.7898 | 0.4344 | 0.5813 | <u>0.1703</u> | 0.8383 | <u>0.2507</u> | <u>0.2633</u> | 0.3748 | 0.1180 | 0.3435 | **0.0005** |
| 2015 | >2 | CCPM | 0.7906 | 0.4349 | **0.6401** | 0.1097 | 0.8383 | 0.2166 | 0.1872 | 0.3748 | 0.1181 | 0.3437 | 0.0834 |
| 2016 | >2 | XGBoost | 0.7417 | 0.3650 | 0.5972 | 0.0636 | 0.8337 | 0.1545 | 0.1149 | 0.4008 | 0.1255 | 0.3542 | 0.0048 |
| 2016 | >2 | PNN | **0.7960** | **0.4441** | 0.6157 | 0.1502 | **0.8398** | 0.2468 | 0.2414 | **0.3702** | **0.1167** | **0.3416** | 0.0166 |
| 2016 | >2 | Wide&Deep | 0.7916 | 0.4375 | 0.6184 | 0.1360 | 0.8390 | 0.2353 | 0.2230 | 0.3731 | 0.1175 | 0.3428 | 0.0128 |
| 2017 | 2 | AFM | 0.7782 | 0.4129 | 0.5887 | 0.1204 | 0.8364 | 0.2116 | 0.1999 | 0.3817 | 0.1201 | 0.3466 | 0.0027 |
| 2017 | >2 | NFM | 0.7886 | 0.4320 | <u>0.6235</u> | 0.1237 | 0.8385 | 0.2255 | 0.2063 | 0.3748 | 0.1181 | 0.3437 | 0.0169 |
| 2017 | >2 | DeepFM | <u>0.7951</u> | <u>0.4428</u> | 0.6169 | 0.1470 | <u>0.8396</u> | 0.2445 | 0.2374 | <u>0.3712</u> | <u>0.1169</u> | <u>0.3420</u> | 0.0306 |
| 2019 | >2 | FiGNN | 0.7896 | 0.4338 | 0.6113 | 0.1389 | 0.8388 | 0.2355 | 0.2263 | 0.3743 | 0.1179 | 0.3434 | <u>0.0022</u> |
| 2019 | >2 | AutoInt | 0.7897 | 0.4345 | 0.6113 | 0.1413 | 0.8389 | 0.2377 | 0.2296 | 0.3747 | 0.1180 | 0.3435 | 0.0575 |
| 2021 | 2 | FmFM | 0.7909 | 0.4346 | 0.5822 | **0.1719** | 0.8384 | **0.2523** | **0.2653** | 0.3747 | 0.1180 | 0.3435 | 0.0038 |
| 2023 | >2 | GDCN | 0.7874 | 0.4325 | 0.6114 | 0.1383 | 0.8388 | 0.2351 | 0.2256 | 0.3754 | 0.1182 | 0.3438 | 0.0145 |

Note: The best results across each metric column are **bolded**, the second-best results are <u>underlined</u>.

Table 6. Model performance comparison on AntM$^2$C

| Year | Feature interaction order | Model | Confusion matrix-based metrics | | | | | | Probability-based metrics | | | |
|------|------|------|------|------|------|------|------|------|------|------|------|------|
| | | | AUC-ROC | AUC-PR | Precision | Recall | Accuracy | MCC | F1 | Logloss | MSE | RMSE | \|1-COPC\| |
| <2000 | 1 | LR | 0.7516 | 0.7370 | 0.7758 | 0.4443 | 0.6692 | 0.3619 | 0.5650 | 0.6391 | 0.2235 | 0.4728 | 0.0625 |
| <2000 | >2 | LSTM | 0.8378 | 0.8329 | 0.7674 | 0.7117 | 0.7563 | 0.5122 | 0.7385 | 0.5119 | 0.1666 | 0.4082 | 0.0778 |
| 2010 | 2 | FM | 0.8383 | 0.8329 | <u>0.8815</u> | 0.5203 | 0.7342 | 0.5033 | 0.6544 | 0.5217 | 0.1746 | 0.4178 | 0.2553 |
| 2015 | >2 | CCPM | 0.8513 | 0.8419 | 0.7288 | **0.8087** | 0.7619 | 0.5279 | **0.7666** | 0.4895 | 0.1610 | 0.4012 | 0.0929 |
| 2016 | >2 | XGBoost | 0.7364 | 0.7714 | **0.9203** | 0.4195 | 0.7017 | 0.4647 | 0.5763 | 0.6375 | 0.2099 | 0.4582 | 0.2859 |
| 2016 | >2 | PNN | 0.8532 | 0.8464 | 0.7440 | <u>0.7882</u> | 0.7665 | 0.5342 | <u>0.7655</u> | 0.4844 | 0.1584 | 0.3980 | 0.0132 |
| 2016 | >2 | Wide&Deep | <u>0.8588</u> | <u>0.8515</u> | 0.7645 | 0.7633 | <u>0.7719</u> | <u>0.5432</u> | 0.7639 | 0.4794 | <u>0.1549</u> | <u>0.3936</u> | 0.0505 |
| 2017 | 2 | AFM | 0.8528 | 0.8431 | 0.7887 | 0.7124 | 0.7686 | 0.5377 | 0.7486 | 0.4853 | 0.1587 | 0.3984 | 0.0652 |
| 2017 | >2 | NFM | 0.8520 | 0.8408 | 0.7884 | 0.7075 | 0.7667 | 0.5341 | 0.7458 | 0.4857 | 0.1587 | 0.3984 | 0.0711 |
| 2017 | >2 | DeepFM | 0.8554 | 0.8484 | 0.7496 | 0.7769 | 0.7666 | 0.5336 | 0.7630 | **0.4724** | 0.1552 | 0.3939 | 0.0532 |
| 2019 | >2 | FiGNN | 0.8500 | 0.8400 | 0.7511 | 0.7739 | 0.7667 | 0.5335 | 0.7623 | 0.4839 | 0.1584 | 0.3980 | 0.0293 |
| 2019 | >2 | AutoInt | 0.8534 | 0.8430 | 0.7711 | 0.7424 | 0.7689 | 0.5371 | 0.7565 | 0.4792 | 0.1568 | 0.3960 | 0.0230 |
| 2021 | 2 | FmFM | 0.8474 | 0.8385 | 0.7686 | 0.7355 | 0.7650 | 0.5293 | 0.7517 | 0.4842 | 0.1587 | 0.3984 | <u>0.0079</u> |
| 2023 | >2 | GDCN | **0.8605** | **0.8518** | 0.7810 | 0.7385 | **0.7727** | **0.5459** | 0.7582 | <u>0.4771</u> | **0.1539** | **0.3923** | 0.0563 |
| 2023 | - | Uni-CTR | 0.8191 | 0.8071 | 0.7578 | 0.6269 | 0.7250 | 0.4517 | 0.6862 | 0.5097 | 0.1709 | 0.4134 | 0.0467 |
| 2024 | - | Qwen2.5-1.5B | 0.4570 | 0.4511 | 0.4350 | 0.2972 | 0.4722 | -0.0723 | 0.3435 | 0.7043 | 0.2555 | 0.5055 | **0.0024** |

Note: The best results across each metric column are **bolded**, the second-best results are <u>underlined</u>.

Table 7. Model performance comparison on the rule-based synthetic dataset

| Year | Feature interaction order | Model | Confusion matrix-based metrics | | | | | | | Probability-based metrics | | | |
|---|---|---|---|---|---|---|---|---|---|---|---|---|---|
| | | | AUC-ROC | AUC-PR | Precision | Recall | Accuracy | MCC | F1 | Logloss | MSE | RMSE | \|1-COPC\| |
| <2000 | 1 | LR | 0.9045 | 0.7709 | **0.7104** | 0.6052 | 0.8396 | 0.5532 | 0.6536 | 0.3300 | 0.1061 | 0.3257 | <u>0.0002</u> |
| <2000 | >2 | LSTM | <u>0.9063</u> | 0.7753 | 0.6745 | 0.7064 | 0.8413 | 0.5838 | 0.6900 | 0.3153 | 0.1020 | 0.3193 | 0.0152 |
| 2010 | 2 | FM | 0.9055 | 0.7733 | <u>0.6968</u> | 0.6332 | 0.8394 | 0.5594 | 0.6635 | 0.3213 | 0.1033 | 0.3214 | 0.0011 |
| 2015 | >2 | CCPM | <u>0.9063</u> | 0.7753 | 0.6750 | 0.7036 | 0.8412 | 0.5826 | 0.6890 | 0.3151 | <u>0.1019</u> | <u>0.3192</u> | 0.0010 |
| 2016 | >2 | XGBoost | 0.9050 | 0.7722 | 0.6914 | 0.6467 | 0.8395 | 0.5632 | 0.6683 | 0.3221 | 0.1036 | 0.3218 | 0.0007 |
| 2016 | >2 | PNN | **0.9064** | **0.7755** | 0.6719 | 0.7148 | **0.8415** | 0.5865 | 0.6927 | **0.3149** | **0.1018** | **0.3191** | 0.0008 |
| 2016 | >2 | Wide&Deep | **0.9064** | **0.7755** | 0.6716 | <u>0.7162</u> | **0.8415** | <u>0.5870</u> | <u>0.6932</u> | **0.3149** | **0.1018** | **0.3191** | 0.0017 |
| 2017 | 2 | AFM | 0.9053 | 0.7734 | 0.6903 | 0.6559 | 0.8404 | 0.5675 | 0.6726 | 0.3195 | 0.1029 | 0.3208 | 0.0008 |
| 2017 | >2 | NFM | **0.9064** | **0.7755** | 0.6721 | 0.7139 | <u>0.8414</u> | 0.5862 | 0.6924 | <u>0.3150</u> | **0.1018** | **0.3191** | 0.0011 |
| 2017 | >2 | DeepFM | **0.9064** | **0.7755** | 0.6727 | 0.7113 | 0.8413 | 0.5852 | 0.6915 | **0.3149** | **0.1018** | **0.3191** | 0.0004 |
| 2019 | >2 | FiGNN | **0.9064** | <u>0.7754</u> | 0.6711 | **0.7169** | <u>0.8414</u> | **0.5871** | **0.6933** | <u>0.3150</u> | <u>0.1019</u> | **0.3191** | **0.0001** |
| 2019 | >2 | AutoInt | **0.9064** | **0.7755** | 0.6715 | 0.7160 | <u>0.8414</u> | 0.5868 | 0.6930 | **0.3149** | **0.1018** | **0.3191** | 0.0005 |
| 2021 | 2 | FmFM | 0.9055 | 0.7734 | 0.6967 | 0.6335 | 0.8394 | 0.5596 | 0.6636 | 0.3212 | 0.1032 | 0.3213 | 0.0014 |
| 2023 | >2 | GDCN | <u>0.9063</u> | <u>0.7754</u> | 0.6742 | 0.7064 | 0.8413 | 0.5836 | 0.6899 | <u>0.3150</u> | <u>0.1019</u> | <u>0.3192</u> | 0.0008 |

Note: The best results across each metric column are **bolded**, the second-best results are <u>underlined</u>.

Table 8. Model performance evaluation from synthetic to real-world datasets

| Year | Feature interaction order | Model | Confusion matrix-based metrics | | | | | | | Probability-based metrics | | | |
|------|------|------|------|------|------|------|------|------|------|------|------|------|------|
| | | | AUC-ROC | AUC-PR | Precision | Recall | Accuracy | MCC | F1 | Logloss | MSE | RMSE | \|1-COPC\| |
| <2000 | 1 | LR | <u>0.6526</u> | 0.4010 | 0.4422 | 0.3935 | 0.7178 | **0.2319** | 0.4165 | <u>0.5469</u> | <u>0.1818</u> | <u>0.4264</u> | 0.1437 |
| <2000 | >2 | LSTM | 0.6402 | 0.3638 | 0.3446 | <u>0.6106</u> | 0.5781 | 0.1708 | 0.4193 | 0.6001 | 0.2059 | 0.4534 | 0.3328 |
| 2010 | 2 | FM | 0.6438 | 0.4021 | 0.3696 | 0.5455 | 0.6455 | 0.2029 | <u>0.4406</u> | 0.6419 | 0.2144 | 0.4631 | 0.3818 |
| 2015 | >2 | CCPM | 0.6517 | <u>0.4044</u> | 0.4502 | 0.3382 | 0.7250 | <u>0.2172</u> | 0.3862 | 1.2910 | 0.2367 | 0.4866 | 0.3563 |
| 2016 | >2 | XGBoost | 0.6470 | 0.3754 | 0.4421 | 0.2720 | 0.7257 | 0.1842 | 0.3360 | 0.5901 | 0.1935 | 0.4399 | 0.1302 |
| 2016 | >2 | PNN | 0.6479 | 0.3983 | 0.5092 | 0.1413 | 0.7418 | 0.1378 | 0.1913 | 0.5544 | 0.1849 | 0.4300 | <u>0.0329</u> |
| 2016 | >2 | Wide&Deep | 0.6471 | 0.3983 | 0.4378 | 0.3660 | 0.7006 | 0.1989 | 0.3598 | 0.5727 | 0.1925 | 0.4386 | 0.2281 |
| 2017 | 2 | AFM | 0.6503 | 0.4007 | 0.4732 | 0.2624 | 0.7361 | 0.2019 | 0.3364 | **0.5406** | **0.1789** | **0.4230** | **0.0198** |
| 2017 | >2 | NFM | 0.6500 | 0.3974 | **0.5271** | 0.1149 | **0.7465** | 0.1483 | 0.1856 | 0.5771 | 0.1939 | 0.4403 | 0.2049 |
| 2017 | >2 | DeepFM | 0.6487 | 0.3992 | 0.4986 | 0.1698 | 0.7424 | 0.1674 | 0.2458 | 0.5575 | 0.1856 | 0.4308 | 0.1757 |
| 2019 | >2 | FiGNN | 0.6450 | 0.3969 | 0.2416 | 0.0813 | 0.7427 | 0.0799 | 0.1216 | 0.5524 | 0.1836 | 0.4285 | 0.0398 |
| 2019 | >2 | AutoInt | 0.6509 | 0.3978 | <u>0.5160</u> | 0.0563 | <u>0.7454</u> | 0.0974 | 0.0986 | 0.5592 | 0.1863 | 0.4317 | 0.1566 |
| 2021 | 2 | FmFM | **0.6557** | **0.4065** | 0.3721 | 0.5684 | 0.6424 | 0.2120 | **0.4487** | 0.5852 | 0.1979 | 0.4448 | 0.2789 |
| 2023 | >2 | GDCN | 0.6376 | 0.3762 | 0.2830 | **0.8714** | 0.4016 | 0.1187 | 0.4271 | 2.0144 | 0.4560 | 0.6752 | 0.6412 |

Note: The best results across each metric column are **bolded**, the second-best results are <u>underlined</u>.

### 6.3.4 Results analysis

In order to provide intuitive visualizations of model performance across all datasets, we employ radar graphs to present experimental results for the first-order, second-order and high-order feature interaction models. Figures 7(a)-7(d) display the experimental results using two radar graphs for each dataset: the left graph represents model performance in terms of confusion matrix-based metrics, where higher values indicate better performance, and the right graph represents model performance in terms of probability-based metrics, where lower values indicate better performance. Three distinct color families denote models by feature interaction order: the yellow line represents the first-order feature interaction model, the blue lines represent the second-order feature interaction models, and the red lines represent high-order feature interaction models. In order to make relative performance gaps apparent in the radar graphs, each metric value is normalized based on its actual range.

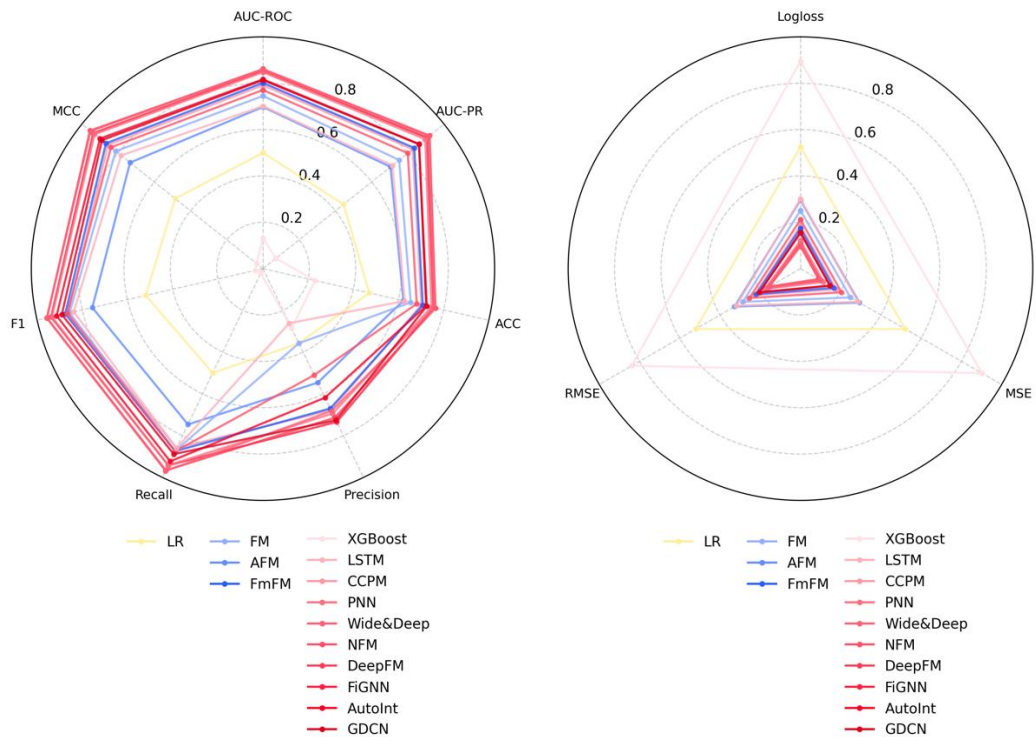

Figure 7(a). Radar graphs of model performance on Criteo

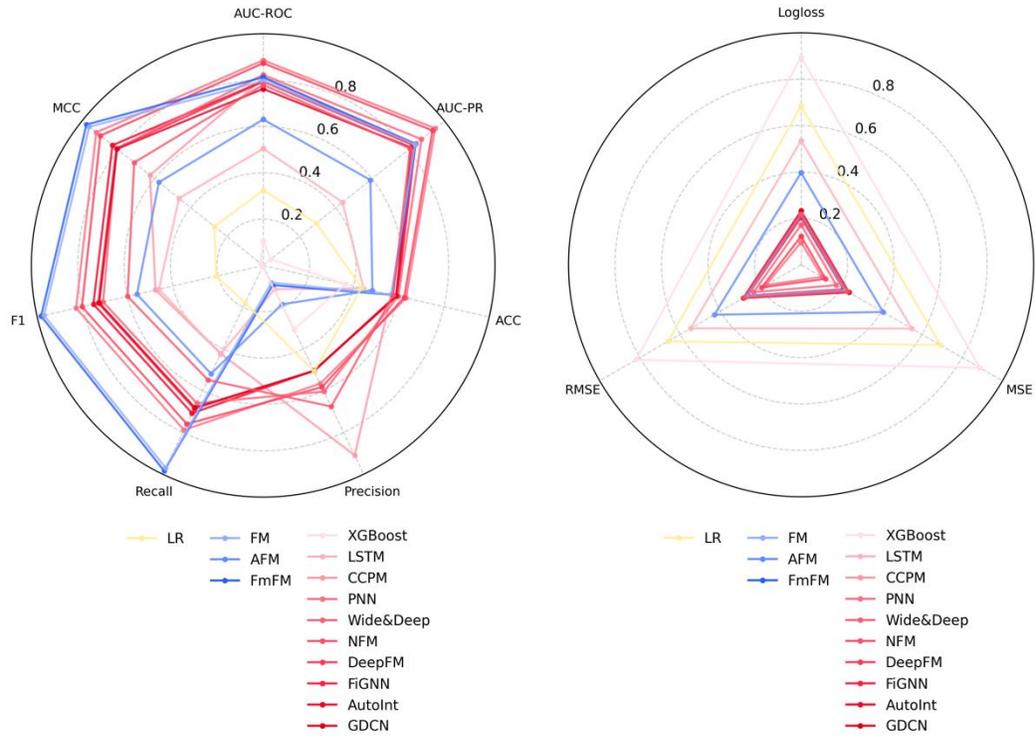

Figure 7(b). Radar graphs of model performance on Avazu

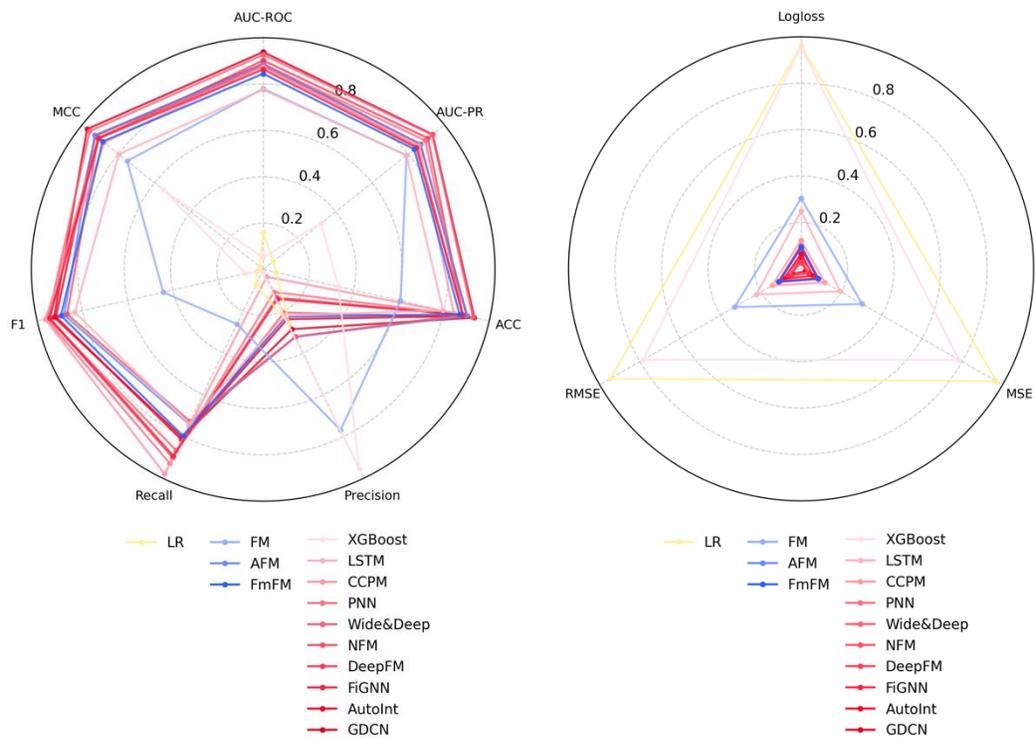

Figure 7(c). Radar graphs of model performance on AntM$^2$C

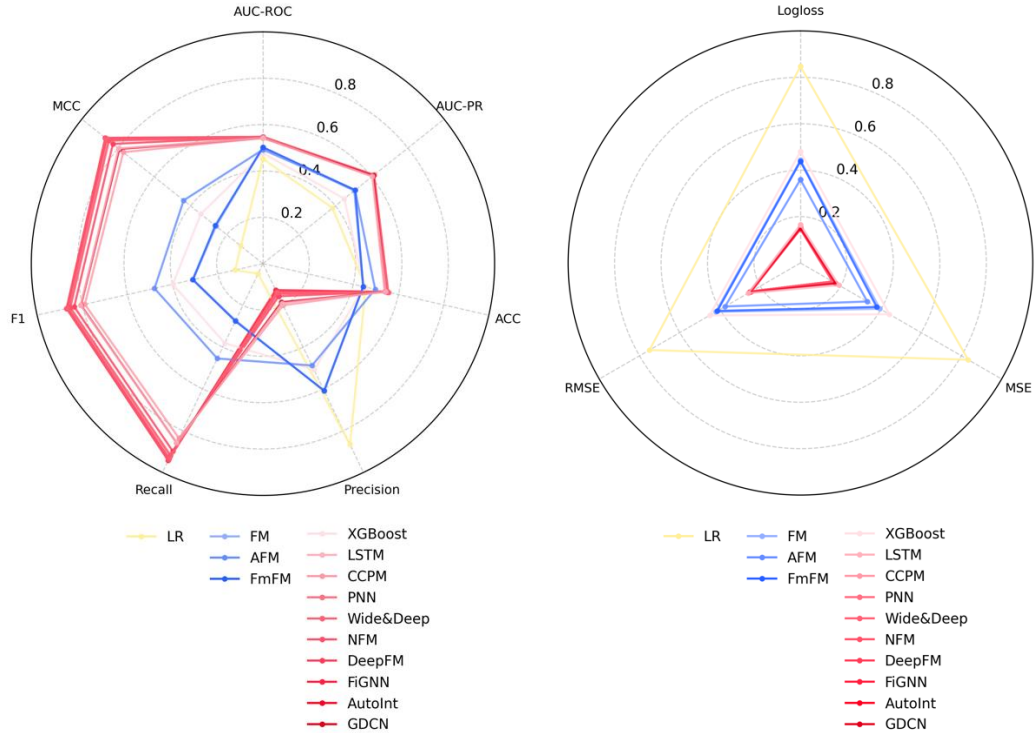

Figure 7(d). Radar graphs of model performance on the rule-based synthetic dataset

From Figures 7(a)-7(d), we can observe the following results.

First, the radar graphs of confusion matrix-based metrics reveal that high-order models (red lines) generally outperform low-order models (blue and yellow lines) with larger areas. Similarly, the radar graphs of probability-based metrics also reveal that high-order models achieve better performance, as the red lines are enclosed within the blue and yellow lines. In other words, high-order models outperform low-order models despite some exceptions.

Second, on Criteo, high-order models generally outperform low-order models in terms of all metrics. However, on Avazu, the second-order models outperform high-order models in terms of some metrics (i.e., MCC, F1 score, and recall). Similarly, on AntM$^2$C and the rule-based synthetic datasets, the second-order models outperform high-order models in terms of precision. This phenomenon demonstrates that model performance is affected by dataset characteristics.

Third, different evaluation metrics show varying sensitivity to high- and low-order models. Specifically, high-order models generally outperform low-order models in terms of AUC-ROC, AUC-PR, and accuracy on all the datasets, suggesting that these metrics provide stable measurements. In contrast, the performance of high- and low-order models shows obvious fluctuation in terms of precision and recall, demonstrating that when selecting these evaluation metrics researchers need to consider model architectures and dataset characteristics.

Next, we visualize the evolution of model performance on three public datasets (i.e., Criteo, Avazu, and AntM$^2$C), as shown in Figure 8. Figure 8 presents four subplots that show the average and best model performance in terms of AUC-ROC and Logloss from 2015 to 2021, where the orange line denotes the results on Avazu, the green line denotes the results on Criteo, and the blue line denotes the results on AntM$^2$C. Notably, we exclude 2023 as the Uni-CTR was fine-tuned with only 2% of the dataset, which would be insufficient for optimal performance and would skew the analysis.

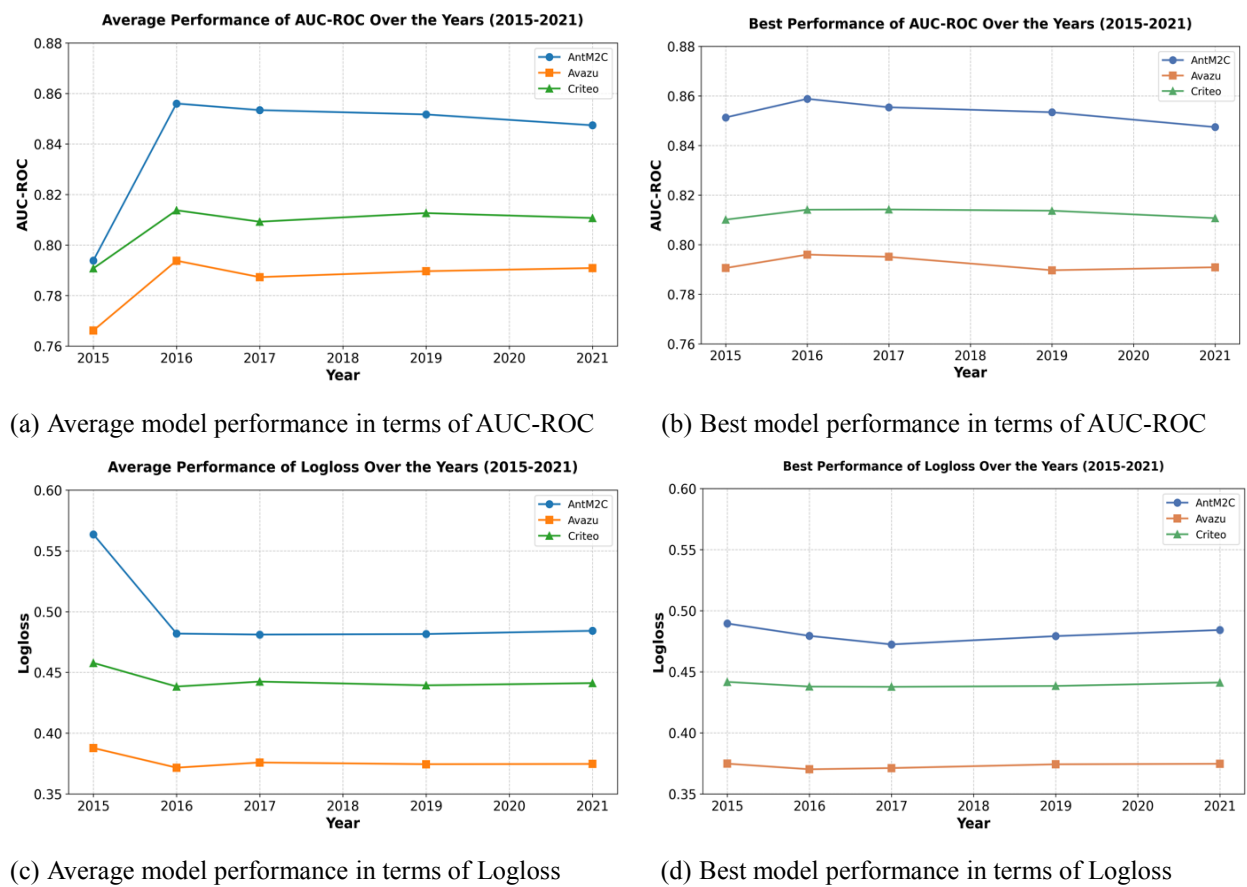

(a) Average model performance in terms of AUC-ROC

(b) Best model performance in terms of AUC-ROC

(c) Average model performance in terms of Logloss

(d) Best model performance in terms of Logloss

Figure 8: Performance evolution of CTR prediction models over years from 2015 to 2021

From Figure 8, we observe the phenomenon that both the average and best model performance in terms of AUC-ROC and Logloss show significant improvements from 2015 to 2016, particularly in average performance. This substantial progress in the early period likely correlates with the introduction of deep learning models. After 2016, however, model performance tends to stabilize with relatively slow progress, indicating a performance bottleneck period in CTR prediction tasks. Notably, the performance gaps across different datasets remain stable over time, indicating that dataset characteristics consistently influence model performance.

# 7. Research perspectives

In this section, we discuss the main challenges and future directions in the area of advertising CTR prediction.

## 7.1 Main challenges

### 7.1.1 Low code availability in CTR prediction

Although a large number of CTR prediction models have been developed, only a small portion of articles are accompanied by code. Even though, code shared by some studies lacks essential components such as data preparation, feature engineering and embedding. The low code availability and code incompleteness raise a significant challenge for reproducing CTR prediction models. Thereby, many models remain unverifiable in comparative studies, which hampers progress in the area of CTR prediction.

To address this issue, the community (e.g., researchers, journals and institutions) should meet an agreement that the transparency of code is beneficial to scientific progress in this area. Meanwhile, it is suggested for the community to implement explicit code publishing policies for encouraging researchers and practitioners to share code and ideas.

### 7.1.2 Lack of benchmarking datasets in CTR prediction

Extant CTR prediction research primarily relies on public datasets such as Criteo and Avazu. These datasets which were released over a decade ago, reflect relatively traditional advertising formats and fail to support emerging advertising formats, e.g., interactive video advertising, virtual reality (VR)-based advertising, and artificial intelligence (AI) generated advertising. Moreover, features in the public datasets might be anonymized due to privacy concerns, which may hamper the evaluation of models relaying on semantics and knowledge graphs and decrease the model interpretation. This situation has become particularly worse with LLM-based CTR prediction models that require multi-modal features and explicit definitions of features.

Large-scale multi-modal datasets are expected to alleviate this predicament (Huan et al., 2024). An alternative way is to develop benchmarking datasets incorporating new digital advertising formats and design synthetic data generation algorithms enabling researchers to conduct controllable experiments.

### 7.1.3 Increasing computational resource demands in CTR prediction

The advanced technologies have improved the performance of CTR prediction models. However, the computational resource demands have increased significantly. This resource consumption challenge is caused by three main factors. First, embedding parameters consume substantial computational space due to sparse features (Shi et al., 2020; Jin & Yang, 2025), leading to inefficient memory utilization. Second, the adoption of deep learning methods increases model architecture complexity. These models are concentrated in the embedding module (Guo et al., 2021), with parameters that may exceed 10TB even after quantization (Zhao et al., 2019). The integration of large language models escalates this computational resource demand, as exemplified by Uni-CTR (Fu et al., 2023) employing a language model with 1.5 billion parameters. Third, the large-scale datasets such as AntM$^2$C (Huan et al., 2024) with 1.1 billion instances require greater computational power for data processing and model training.

Fortunately, several recently developed solutions have shown potential efficacies for addressing this challenge, e.g., distillation methods saving substantial computational resources while maintaining the competitive model performance (Wang et al., 2023a), and centralized training systems employing solid-state drives (SSDs) and GPUs (Zhao et al., 2019) to train CTR prediction models with tens-of-terabytes-scale parameters.

### 7.1.4 Data imbalance in CTR prediction

In online advertising, CTR prediction datasets are highly imbalanced, with the number of non-clicks vastly outnumbering the clicked ones. This imbalanced distribution of instances impacts the training and evaluation processes of CTR prediction models (Liu et al., 2018; Liu et al., 2023; Zhang et al., 2024c) in that models tend to excel at predicting the majority class (i.e., non-clicks) while performing poorly on the minority class (i.e., clicks). Recently, researchers have made preliminary investigations on this issue by using techniques such as regularized adversarial sampling and transfer learning (Xie et al., 2019; Liu et al., 2023).

## 7.2 Future directions

### 7.2.1 Incorporating LLMs into CTR prediction models

Large language models (LLMs) demonstrate significant potential for improving CTR prediction through their semantic understanding abilities and extensive world knowledge (Fu et al., 2023; Yuan et al., 2023). Specifically, LLMs can be employed to extract and capture meaningful semantic signals (Xi et al., 2024), enrich semantic feature embeddings (Qiu et al., 2024), and generate summaries for advertisements (Dubey et al., 2024). However, most existing studies primarily treat LLMs as independent components for capturing semantic signals, ignoring the relationships between semantic signals and graph structural signals. A promising direction is to integrate LLMs and graph neural networks (GNNs) into a unified architecture for CTR prediction, where GNNs capture feature interactions among users and advertisements via graph structures and learn representations through message aggregation and propagation, and LLMs process textual information to extract semantic signals that can be incorporated into the graph structure to enhance representation learning. This integration could improve model performance by leveraging semantic signals from text information and structural patterns from graphs.

### 7.2.2 Contrastive learning for multi-modal CTR prediction

Multi-modal content that contains rich information for CTR prediction tasks is available on online advertising platforms, e.g., textual descriptions, product images, and promotional videos. However, many CTR prediction models primarily rely on categorical and numerical features and fail to exploit multi-modal features (He et al., 2021). In recent years, researchers have attempted to leverage multi-modal features in CTR prediction by merging heterogeneous features into weighted vectors (Chen et al., 2020), and decoupling multi-modal feature interaction into independent components (Li et al., 2024a). However, these methods struggle to effectively fuse multi-modal information and learn discriminative representations (Xiao et al., 2022) due to the heterogeneity gaps between different modalities.

Contrastive learning presents a promising approach to bridge the gaps between multi-modal features and enhance the discriminative capability of CTR prediction models. Specifically, models are trained to distinguish between positive and negative pairs, where positive pairs consist of different modalities representing the same entity (i.e., users and advertisements), and negative pairs contain cross-modal combinations from different entities.

### 7.2.3 Multi-task learning for CTR prediction

In the context of online advertising, researchers usually independently process multiple prediction tasks, including CTR prediction and conversion rate (CVR) estimation (Xue et al., 2025). This approach often yields suboptimal performance because different prediction tasks are inherently interconnected (Li et al., 2020a; Yang et al., 2022). Multi-task learning can enhance CTR prediction by capturing these underlying relationships and transferring knowledge among related tasks.

In recent years, there has been a growing trend of introducing specialized frameworks for multi-task learning, e.g., expert-based frameworks (Ma et al., 2018; Tang et al., 2020), neural network-based frameworks (Fei et al., 2021), end-to-end joint optimization frameworks (Yang et al., 2021), and entire-space causal multi-task frameworks (Zhu et al., 2023). Several studies have extended these frameworks to simultaneously process multiple prediction tasks in different advertising environments (Zhang et al., 2022; Min et al., 2023). It is a promising direction to develop frameworks that dynamically optimize knowledge sharing based on task relationships and advertising scenarios for CTR prediction.

### 7.2.4 Causal inference for CTR prediction

Existing CTR prediction models primarily rely on capturing correlations rather than causal relationships, limiting their generalization to unseen data. Causal inference improves the performance of CTR prediction models by learning the cause-and-effect relationships in users' behaviors (Gao et al., 2024).

In this direction, causal inference has been employed to capture high-order interactions (Zhai et al., 2023) and improve result interpretability (Sui et al., 2022). Further exploration of causal inference for CTR prediction is a promising direction. Specifically, causal discovery methods can be employed to identify key variables that drive user clicks, thereby guiding model design and feature selection.

## 8. Conclusion

In this paper, we provide a benchmark for CTR prediction in the context of online advertising. First, we propose a unified architecture of CTR prediction benchmark (Bench-CTR) platform that offers flexible interfaces with datasets and components of a wide range of CTR prediction models. Second, we construct a comprehensive system of evaluation protocols encompassing real-world and synthetic datasets, a taxonomy of metrics, standardized procedures and experimental guidelines for performance assessment of CTR prediction models. Third, we implement Bench-CTR and conduct a comparative study to validate 15 CTR prediction models in terms of 11 metrics on three public datasets and two synthetic datasets. Results reveal that, (1) high-order models largely outperform low-order models, though such advantage varies in terms of metrics and on different datasets; (2) models based on large language models demonstrate a remarkable data efficiency, i.e., achieving comparable performance to other models while using only 2% of the training data; (3) the performance of CTR prediction models has achieved significant improvements from 2015 to 2016, then reached a stage with slow progress, which is consistent across various datasets. This benchmark serves as an environment for model development and evaluation, and enhances practitioners' understanding of the underlying mechanisms of models in the area of CTR prediction.

In future research, we plan to explore domain-specific metrics (e.g., return on investment) for advertising effectiveness by conducting field research or simulating market mechanisms and processes in online advertising environments. Second, we will continue to incorporate newly developed models into the Bench-CTR platform to keep pace with the development of CTR prediction research. Last but not least, it is interesting to explore the role of LLMs in CTR prediction evaluation by leveraging their remarkable capabilities in semantic understanding. LLMs can be utilized to generate summaries by analyzing comparative results of various models, which helps to improve the interpretability of prediction results.

## Declarations

Funding and/or Conflicts of interests/Competing interests: This work is partially supported by the (NSFC National Natural Science Foundation of China) grants (72171093). The authors have no competing interests to declare.

# Appendix for "Toward a benchmark for CTR prediction in online advertising: datasets, evaluation protocols and perspectives"

Shan Gao[1], Yanwu Yang[1]

[1]School of Management, Huazhong University of Science and Technology, Wuhan 43004, China,
{gaoshan.isec, yangyanwu.isec}@gmail.com

## A.1 List of abbreviations

Table A1. List of abbreviations used in this paper.

| Abbreviations | Full term |
| --- | --- |
| ACN | Attentive Capsule Network |
| AFM | Attentional Factorization Machine |
| AFN | Adaptive Factorization Network |
| AI | Artificial Intelligence |
| ARC | AI2 Reasoning Challenge |
| AUC-PR | Area Under Precision Recall Curve |
| AUC-ROC | Area Under ROC Curve |
| AutoInt | Automatic Feature Interaction Learning |
| BAHE | Behavior Aggregated Hierarchical Encoding |
| CAGR | Compound Annual Growth Rate |
| CCPM | Convolutional Click Prediction Model |
| CIFAR | Canadian Institute for Advanced Research |
| CIN | Compressed Interaction Network |
| CNN | Convolutional Neural Network |
| COPC | Click Over Predicted Click |
| CTR | Click-Through Rate |
| CVR | Conversion Rate |
| DCN | Deep&Cross Network |
| DeepFM | Factorization-Machine based Neural Network |
| DeepGBM | Deep Learning Framework Distilled by GBDT |
| DIEN | Deep Interest Evolution Network |
| DMCNN | Density Matrix-Based Convolutional Neural Network |
| DNN | Deep Neural Network |
| DSP | Demand-Side Platform |
| EDA | Exploratory Data Analysis |
| ETCF | Ensemble Trees and Cascading Forests |
| FFM | Field-aware Factorization Machine |
| FiBiNET | Feature Importance and Bilinear feature Interaction NETwork |
| Field-ECE | Field-level Expected Calibration Error |
| Field-RCE | Field-level Relative Calibration Error |
| FiGNN | Feature Interaction Graph Neural Networks |
| FinalMLP | Feature Selection and Interaction Aggregation Layers on Top of Two MLP |
| FM | Factorization Machine |
| FmFM | Field-matrixed Factorization Machine |
| FN | False Negative |
| FP | False Positive |
| FPR | False Positive Rate |
| FwFM | Field-weighted Factorization Machine |
| GBDT | Gradient Boosting Decision Tree |
| gcForest | Multi-grained Cascade Forest |
| GCN | Gated Cross Network |
| GDCN | Gated Deep Cross Network |
| GLUE | General Language Understanding Evaluation |
| GNN | Graph Neural Network |
| GraphFM | Graph Factorization Machine |
| HCCM | Hybrid CNN based Attention with Category Prior Module |
| HellaSwag | Harder Endings, Longer contexts, and Low-shot Activities for Situations With Adversarial Generations |
| HoAFM | High-order Attentive Factorization Machine |
| InterHAt | Interpretable CTR prediction model with Hierarchical Attention |
| KLD | Kullback-Leibler Divergence |
| LLM | Large Language Model |
| LoRA | Low-Rank Adaptation |
| LR | Logistic Regression |
| LSTM | Long Short-Term Memory |
| MCC | Matthews Correlation Coefficient |
| MLP | Multi-Layer Perceptron |
| MLLM | Multi-modal Large Language Model |
| MSE | Mean Square Error |
| MT-bench | Multi-Turn Benchmark |
| NFM | Neural Factorization Machine |
| ONN | Operation-aware Neural Networks |
| OOV | Out-of-Vocabulary |
| PNN | Product-based Neural Network |
| Poly2 | Degree-2 Polynomial Model |
| RelaImpr | Relative Improvement |
| RIG | Relative Information Gain |
| RL | Reinforcement Learning |
| RMSE | Root Means Squared Error |
| RNN | Recurrent Neural Network |
| ROC Curve | Receiver Operating Characteristic Curve |
| SENET | Squeeze-Excitation network |
| SMOTE | Synthetic Minority Over-sampling Technique |
| SSDs | Solid-State Drives |
| TN | True Negative |
| TP | True Positive |
| TPR | True Positive Rate |
| TVAE | Tabular Variational AutoEncoder |
| Uni-CTR | Unified Framework for Multi-Domain CTR Prediction |
| VR | Virtual Reality |
| xDeepFM | eXtreme Deep Factorization Machine |
| XGBoost | eXtreme Gradient Boosting |

## A.2 Statistics on evaluation metric usage in CTR prediction literature

Table A2. The number of articles using each evaluation metric per year in CTR prediction.

| Year | Precision | Recall | Accuracy | MCC | AUC-ROC | AUC-PR | F1 score | RelaImpr | Logloss | MSE | COPC | KLD | Field-ECE | Field-RCE | RMSE | RIG | Total articles |
|------|-----------|--------|----------|-----|---------|--------|----------|----------|---------|-----|------|-----|-----------|-----------|------|-----|----------------|
| 2012 | 0 | 0 | 0 | 0 | 2 | 0 | 0 | 0 | 1 | 2 | 0 | 0 | 0 | 0 | 0 | 0 | 4 |
| 2013 | 0 | 0 | 0 | 0 | 2 | 0 | 0 | 0 | 0 | 0 | 0 | 0 | 0 | 0 | 0 | 0 | 2 |
| 2014 | 0 | 0 | 0 | 0 | 2 | 1 | 0 | 0 | 1 | 0 | 0 | 0 | 0 | 0 | 1 | 2 | 4 |
| 2015 | 0 | 0 | 0 | 0 | 2 | 0 | 0 | 0 | 1 | 0 | 0 | 0 | 0 | 0 | 0 | 0 | 3 |
| 2016 | 1 | 1 | 0 | 0 | 9 | 0 | 1 | 0 | 6 | 0 | 0 | 0 | 0 | 0 | 4 | 1 | 12 |
| 2017 | 0 | 0 | 0 | 0 | 9 | 0 | 0 | 0 | 6 | 0 | 0 | 0 | 0 | 0 | 0 | 1 | 11 |
| 2018 | 1 | 1 | 1 | 0 | 13 | 0 | 1 | 2 | 7 | 0 | 0 | 0 | 0 | 0 | 0 | 1 | 14 |
| 2019 | 0 | 0 | 2 | 0 | 21 | 0 | 0 | 0 | 14 | 1 | 0 | 1 | 0 | 0 | 1 | 0 | 22 |
| 2020 | 1 | 1 | 1 | 0 | 29 | 0 | 1 | 1 | 23 | 1 | 0 | 0 | 1 | 1 | 5 | 0 | 33 |
| 2021 | 0 | 0 | 2 | 1 | 12 | 0 | 0 | 1 | 10 | 0 | 0 | 1 | 0 | 0 | 0 | 0 | 16 |
| 2022 | 0 | 0 | 0 | 1 | 14 | 1 | 1 | 1 | 7 | 0 | 2 | 0 | 0 | 0 | 0 | 0 | 16 |
| 2023 | 0 | 0 | 0 | 0 | 19 | 0 | 0 | 4 | 13 | 0 | 0 | 0 | 1 | 0 | 0 | 2 | 19 |
| 2024 | 0 | 0 | 1 | 0 | 8 | 0 | 0 | 0 | 5 | 0 | 1 | 0 | 0 | 0 | 0 | 0 | 9 |

Table A3. The percentage of articles using each evaluation metric per year in CTR prediction.

| Year | Precision | Recall | Accuracy | MCC | AUC-ROC | AUC-PR | F1 score | RelaImpr | Logloss | MSE | COPC | KLD | Field-ECE | Field-RCE | RMSE | RIG | Total articles |
|------|-----------|--------|----------|-----|---------|--------|----------|----------|---------|-----|------|-----|-----------|-----------|------|-----|----------------|
| 2012 | 0 | 0 | 0 | 0 | 0.50 | 0 | 0 | 0 | 0.25 | 0.50 | 0 | 0 | 0 | 0 | 0 | 0 | 4 |
| 2013 | 0 | 0 | 0 | 0 | 1.0 | 0 | 0 | 0 | 0 | 0 | 0 | 0 | 0 | 0 | 0 | 0 | 2 |
| 2014 | 0 | 0 | 0 | 0 | 0.50 | 0.25 | 0 | 0 | 0.25 | 0 | 0 | 0 | 0 | 0 | 0.25 | 0.50 | 4 |
| 2015 | 0 | 0 | 0 | 0 | 0.67 | 0 | 0 | 0 | 0.33 | 0 | 0 | 0 | 0 | 0 | 0 | 0 | 3 |
| 2016 | 0.08 | 0.08 | 0 | 0 | 0.75 | 0 | 0.08 | 0 | 0.50 | 0 | 0 | 0 | 0 | 0 | 0.33 | 0.08 | 12 |
| 2017 | 0 | 0 | 0 | 0 | 0.82 | 0 | 0 | 0 | 0.55 | 0 | 0 | 0 | 0 | 0 | 0 | 0.09 | 11 |
| 2018 | 0.07 | 0.07 | 0.07 | 0 | 0.93 | 0 | 0.07 | 0.14 | 0.50 | 0 | 0 | 0 | 0 | 0 | 0 | 0.07 | 14 |
| 2019 | 0 | 0 | 0.09 | 0 | 0.95 | 0 | 0 | 0 | 0.64 | 0.05 | 0 | 0.05 | 0 | 0 | 0.05 | 0 | 22 |
| 2020 | 0.03 | 0.03 | 0.03 | 0 | 0.88 | 0 | 0.03 | 0.03 | 0.70 | 0.03 | 0 | 0 | 0.03 | 0.03 | 0.15 | 0 | 33 |
| 2021 | 0 | 0 | 0.12 | 0.06 | 0.75 | 0 | 0 | 0.06 | 0.62 | 0 | 0 | 0.06 | 0 | 0 | 0 | 0 | 16 |
| 2022 | 0 | 0 | 0 | 0.06 | 0.88 | 0.06 | 0.06 | 0.06 | 0.44 | 0 | 0.12 | 0 | 0 | 0 | 0 | 0 | 16 |
| 2023 | 0 | 0 | 0 | 0 | 1 | 0 | 0 | 0.21 | 0.68 | 0 | 0 | 0 | 0.05 | 0 | 0 | 0.11 | 19 |
| 2024 | 0 | 0 | 0.11 | 0 | 0.89 | 0 | 0 | 0 | 0.56 | 0 | 0.11 | 0 | 0 | 0 | 0 | 0 | 9 |